\documentclass[a4paper,10pt]{article}
\usepackage[utf8]{inputenc}
\usepackage{graphicx}       
\usepackage{longtable} 
\usepackage{color}
\usepackage{amssymb,amsmath}
\usepackage{float}
\usepackage{enumitem}
\usepackage{subfigure}
\usepackage{units}
\usepackage[ruled,vlined]{algorithm2e}
\usepackage[noend]{algpseudocode}

\usepackage{authblk}

\title{A 3D-1D coupled blood flow and oxygen transport model to generate microvascular networks}
\author{T. K\"oppl$^1$, E. Vidotto$^1$, B. Wohlmuth$^1$}

\date{\vspace{-5ex}}

\begin{document}
	
\maketitle

\footnotetext[1]{Department of Mathematics, University of Technology Munich, 
	Boltzmannstr. 3, 85748 Garching bei M\"unchen, Germany, koepplto@ma.tum.de, vidotto@ma.tum.de, wohlmuth@ma.tum.de}

\textbf{Key words: blood flow simulations, oxygen transport, \\
	    dimensionally reduced models, vascular growth, flows in \\
	    porous media, 3D-1D coupled flow models}  
\begin{abstract}
In this work, we introduce an algorithmic approach to generate microvascular networks starting from larger vessels that can be reconstructed without noticeable segmentation errors. Contrary to larger vessels, the reconstruction of fine-scale components of microvascular networks shows significant segmentation errors, and an accurate mapping is time and cost intense. Thus there is a need for fast and reliable reconstruction algorithms yielding  surrogate networks having similar stochastic properties as the original ones. The microvascular networks are constructed in a marching way by adding vessels to the outlets of the vascular tree from the previous step. To optimise the structure of the vascular trees, we use Murray's law  to determine the radii of the vessels and bifurcation angles. In each step, we compute the local gradient of the partial pressure of oxygen and adapt the orientation of the new vessels to this gradient. At the same time, we use the partial pressure of oxygen to check whether the considered tissue block is supplied sufficiently with oxygen. Computing the partial pressure of oxygen, we use a 3D-1D coupled model for blood flow and oxygen transport. To decrease the complexity of a fully coupled 3D model, we reduce the blood vessel network to a 1D graph structure and use a bi-directional coupling with the tissue which is described by a 3D homogeneous porous medium. The resulting surrogate networks are analysed with respect to morphological and physiological aspects. 
\end{abstract}

\section{Introduction}

The reconstruction of vascular networks from medical images plays an important role in different research areas and has several application fields. Modelling and simulation of blood flow and transport processes within vascular networks is of great interest, since it provides the possiblity to study the distribution of injected medications \cite{d2007multiscale,dewhirst2017transport}. Moreover, a reliable and robust blood flow model allows one to investigate the impact of diseases and therapeutic procedures in a non-invasive way. Numerical simulation techniques have become an important part in testing hypotheses when experimental investigations are not possible or are limited by accessibility, size and resolution \cite{drzisga2016numerical,el2018investigating,koppl2014influence,koeppl2018numerical,
	nabil2016computational,possenti2019computational,shipley2010multiscale}. 
To be able to perform conclusive blood flow simulations, it is crucial to have a precise description of the network. This means that accurate data on the radii, lengths and connectivity of the vessels as well as the location of the vessels are required. In order to obtain high-quality data, remarkable progress using high-resolution imaging techniques has been made. As a result, the acquisition of suitable image data with respect to complex vascular systems on different scales has been facilitated \cite{reichold2011cerebral}. 
Nevertheless, an accurate reconstruction of vascular networks is still challenging, in particular, if fine-scale microvascular networks are considered. Most algorithms are concentrated on vessel detection and segmentation, while the determination of the vascular connectivity is neglected \cite{rempfler2014extracting}. Further problems are associated with the image recording (image noise and artifacts) and sample preparation (air bubbles and clotting) \cite{lesage2009review}.

Motivated by such problems, different approaches for generating microvascular networks have been developed \cite{dankelman2007relation}. One approach considers the theory of porous media. Thereby, the fine scaled vessels are homogenised and considered as a porous media. The pores and pore throats within these porous media are given by the blood vessels and its branchings \cite{vidotto2019hybrid}. In order to account for the vessel hierarchy multi-compartment flow models have been considered \cite{hyde2013parameterisation,rohan2018modeling}. A challenge in this context is to estimate the permeability tensors and porosities for each compartment and to derive suitable coupling conditions between the different compartments. Other types of models directly simulate vascular growth taking optimality principles like the minimisaton of building material or the minimisation of energy dissipation into account \cite{lloyd2010optimization,nekka1996model,pries2014making}. Instead of simulating the growth of the vessels, a direct optimisation of the microvascular system can be considered. In this case, the optimisation processes are concentrated on intravascular volume minimisation with constraints based on physiological principles \cite{georg2010global}.

In this work, we adapt concepts developed in \cite{schneider2014tgif,schneider2012tissue} to generate artificial arterial trees. Thereby the authors use ideas from modelling approaches of angiogenesis processes \cite{pries2001structural,pries2005remodeling,secomb2013angiogenesis}. One of these concepts is based on the assumption that the arterial tree generation is driven by the oxygen distribution in the tissue. After measuring the oxygen concentrations, the intensity of vascular endothelial growth factors (VEGFs) triggering the sprouting of new vessels is computed. Concerning the choice of the vessel radii and bifurcation angles, minimisation principles developed by Murray are used \cite{murray1926physiological2,murray1926physiological}. Furthermore, other algorithms may be applied to optimise the vascular tree e.g. by removing tiny and redundant vessels. The arterial tree is constructed stepwise. In each step, firstly, the oxygen concentration in the tissue is computed. According to the new distribution of oxygen and Murray's princples, the directions of growth for new vessels are determined. Redundant vessels are removed before obtaining the final network structure.

For our modelling approach, we adapt some of the ideas presented in \cite{schneider2012tissue} to simulate the growth of an artificial network. However, we are not only focusing on the generation of an arterial tree. In a first step, we consider arterioles and venoles that are well reconstructed from given image data. Based on these vessels, the missing part of the microvascular network linking the larger vessels is generated.  In this context, we do not restrict ourselves to the arterial component, but our starting network can comprise both venous and arterial components. A further enhancement is related to the computation of the oxygen concentration in a tissue block that is supplied by the microvascular network. According to \cite[Appendix A]{schneider2012tissue}, the oxygen concentration is approximated by a superposition of heuristically chosen functions. In the submitted work, we consider 3D-1D coupled models for simulating blood flow and transport of oxygen in microvascular networks. The term 3D-1D indicates that the microvascular network is described by a one-dimensional (1D) graph-like structure while the surrounding tissue is considered as a three-dimensional (3D) porous medium. This implies that blood flow and the transport of oxygen within the microvascular are governed by 1D PDEs, while flow and oxygen transport in the tissue are modelled by Darcy's law \cite{helmig1997multiphase} and a standard convection-diffusion equation. The convection-diffusion equation is  enhanced by the Michaelis-Menten law modelling the consumption of oxygen by the tissue cells \cite[Chapter 2]{d2007multiscale}. To couple the PDE systems for flow and transport, we use a specific coupling concept presented in \cite{cerroni2019mathematical,koppl2018mathematical,kremheller2019approach}. The key ingredient of this concept is to couple the source terms of the PDEs for flow and transport. Thereby, Starling's filtration law or the Kedem-Katchalsky equation \cite{cattaneo2014fem} are used to determine the exchange of fluids and oxygen. In order to embed the 1D PDEs into the source terms of the 3D PDEs, we use Dirac measures concentrated on the vessel surfaces, instead of other approaches, such as \cite{d2008coupling}. This is motivated by the fact that the actual exchange processes between the vascular system and the tissue take place at these locations.

The remainder of this work is structured as follows: Section \ref{sec:Modelling} is devoted to the modelling assumptions, the mathematical equations for blood flow and oxygen transport, as well as the generation of the blood vessel network. Moreover, we discuss the numerical methods that have been used to solve the mathematical model equations. In Section \ref{sec:SimulationResults}, we show numerical results and report on  the influence of typical parameters as well as statistical characterisations. Finally, in Section \ref{sec:Conclusion}, we summarise our main insights and the key ideas developed in this paper.

\section{Mathematical modelling}
\label{sec:Modelling}

The main section of this paper is divided into six subsections. In the first subsection, 
we present the data set for a microvascular network that is used for our numerical simulations in Section \ref{sec:SimulationResults}. Next, the basic modelling assumptions that are used in the following subsections are listed and motivated. To simulate blood flow and transport processes within a microvascular network, we develop in the third and fourth subsection dimension-reduced models for flow and transport in the network and the surrounding tissue. Thereby, flow and transport in the vascular system are governed by 1D models, while for flow and transport in tissue 3D porous media models are considered, as noted earlier. Concerning the transport processes, we restrict ourselves 
to the transport of oxygen. This is motivated by the fact that a suitable microvascular network has to prevent an unbalanced distribution of the partial pressure of oxygen $\left(PO_2 \right)$ within the tissue block under consideration. Based on the $PO_2$ distribution in the tissue, we develop in the next subsection a model for artificial vascular growth to produce a network able to supply the tissue block with a sufficient amount of oxygen. In the final subsection, we briefly describe the numerical methods, used to solve the equations governing the mathematical model.

\begin{figure}[h!]
	\begin{center}		
		\includegraphics[width=0.48\textwidth]{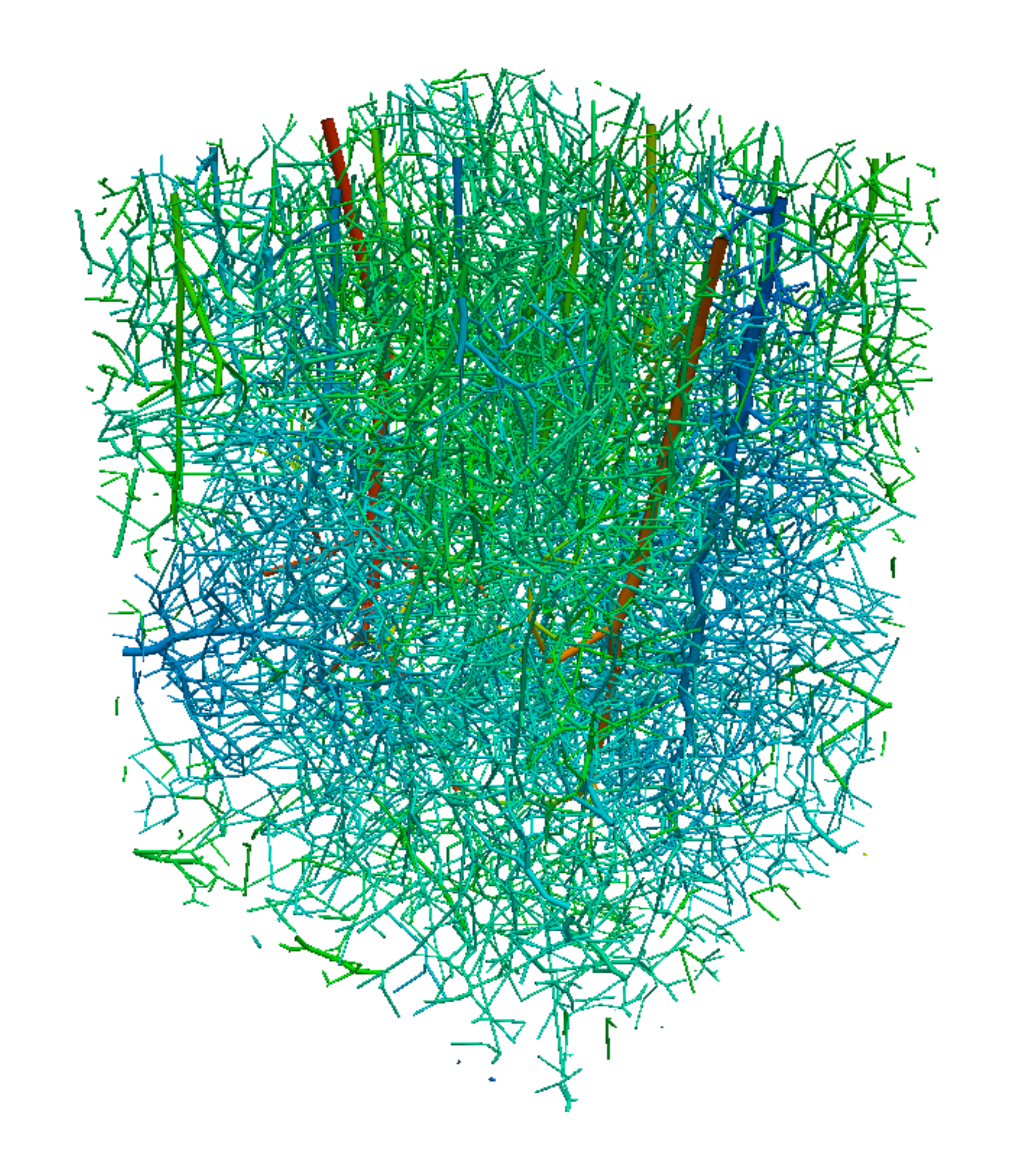}
		\includegraphics[width=0.45\textwidth]{large_vessels.pdf}
	\end{center}
	\caption{\label{fig:networks} Microvascular network extracted from a rat brain contained in a
		volume of approximately $1.0\;\unit{mm} \times 1.0\;\unit{mm} \times 1.5\;\unit{mm}$ (left). Larger arteries and veins contained in this vascular network (right).}
\end{figure}

\begin{figure}[h!]
	\begin{center}		
		\includegraphics[width=1.0\textwidth]{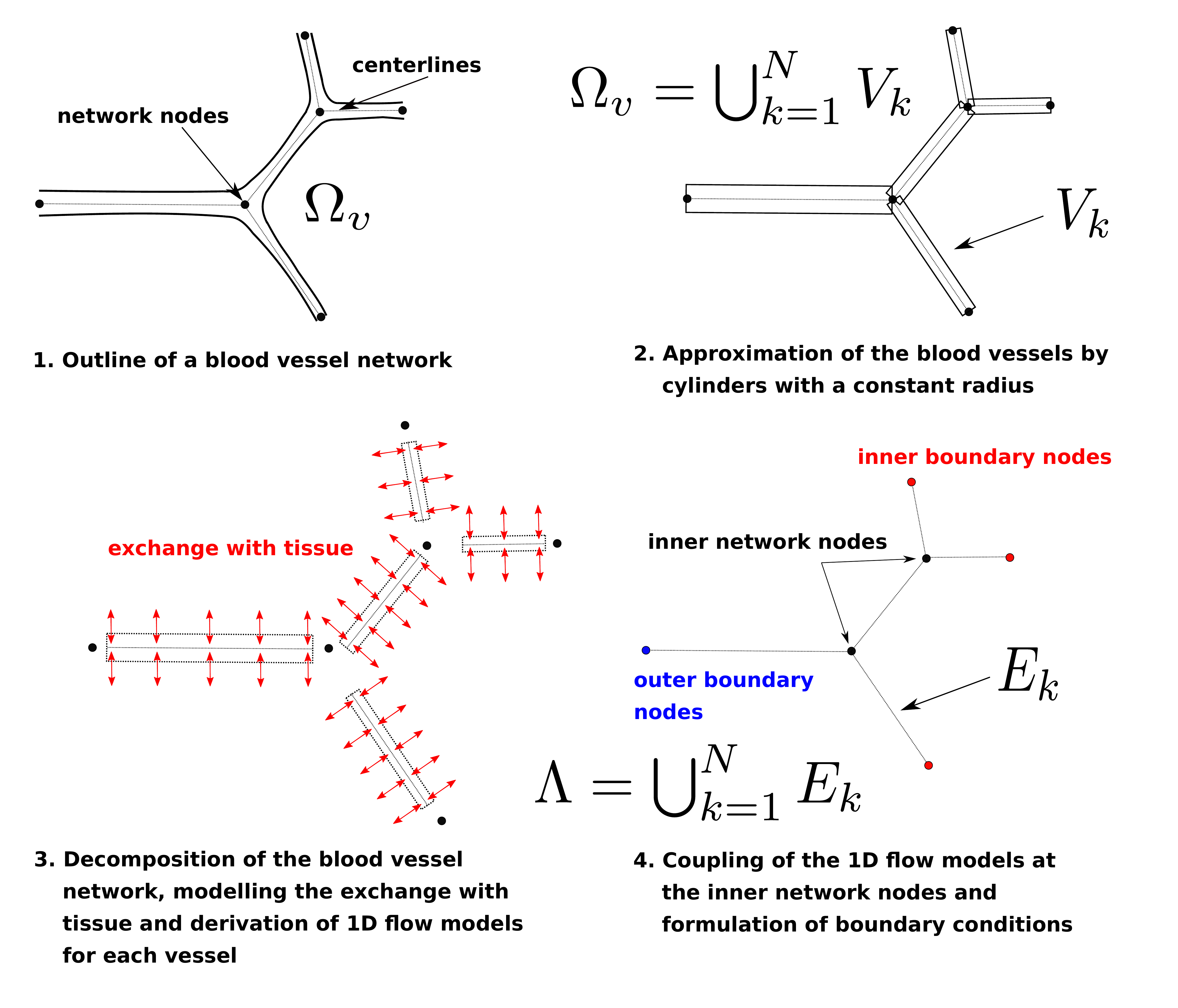}
	\end{center}
	\caption{\label{fig:DD} This figure illustrates the four basic steps for modelling of blood flow and oxygen transport in microvascular networks. In a first step, we consider the outline of a microvascular network $\Omega_v$. Next the blood vessels are approximated by straight cylinders. After that, the network is decomposed, and 1D models for blood flow and oxygen transport for each vessel are derived. In the last step, the 1D models are coupled and boundary conditions are derived, such that blood flow and oxygen transport can be simulated within the whole 1D network.}
\end{figure}

For our numerical simulations, we use the microvascular network shown in Figure \ref{fig:networks} (left). This microvascular network is part of a larger vascular network taken from the cortex of a rat brain. The whole vascular network is described in \cite{reichold2009vascular}. According to this reference, its vessel midlines together with the 3D coordinates of branches, kinks as well as inlets and outlets 
are reconstructed from X-ray images. In the remainder of this paper, we denote these points as network nodes. 
Curved midlines are approximated by straight edges and for each vessel a mean radius is estimated. To each network node that is located within the sample, a fixed pressure value is assigned. 
The larger vessels of this network considered in this work are contained in a domain $\Omega_{roi} \subset \mathbb{R}^3$ given by (roi: region of interest):
\begin{align*}
\Omega_{roi} = \left\{ \mathbf{x} \in \mathbb{R}^3 \left|\; L_{1,l}^r < x_1 < L_{1,u}^r,\; L_{2,l}^r < x_2 < L_{2,u}^r,\;L_{3,l}^r < x_3 <  L_{3,u}^r \right. \right\}.
\end{align*}
However, we consider a larger cuboid $\Omega,\;\Omega_{roi} \subset \Omega \subset \mathbb{R}^3$ as our tissue domain:
\begin{align*}
\Omega = \Big\{\mathbf{x} \in \mathbb{R}^3 \left| \;L_{1,l} < x_1 <  L_{1,u},\;
L_{2,l} < x_2 < L_{2,u},\; L_{3,l} < x_3 <  L_{3,u} \right. \Big\}.
\end{align*} 
Thereby, we choose $\Omega$ such that $\Omega_{roi} \subset \Omega$. This is motivated by the fact that if we consider $\Omega_{roi}$ as a tissue domain, the resulting artificial network might be affected by the chosen boundary conditions. In order to determine the larger domain $\Omega$, each edge of $\Omega_{roi}$ is enlarged on both ends by about $10\;\%$. The values for the lengths defining $\Omega_{roi}$ are provided in Table \ref{tab:model_parameters}. We assume further that the given microvascular network can be represented by a union of $N$ cylinders $V_k$ with a radius $R_k$ and a straight center line or edge $E_k,$ $k\in \left\{1,\ldots,N\right\}$, see Step 2 in Figure \ref{fig:DD}. By this, the domain $\Omega$ can be decomposed in two parts:
$$
\Omega_{v}=\bigcup_{i=1}^N V_k\qquad\text{and}\qquad \Omega_{t}=\Omega\setminus \Omega_{v},
$$
where $\Omega_{v}$ stands for the vascular system and $\Omega_{t}$ the tissue. We denote the two endpoints of $\Lambda_k$ by $\mathbf{x}_{k,1},\mathbf{x}_{k,2}\in\Omega$ and the length of $V_k$ by $l_k =\left\| \mathbf{x}_{k,2}-\mathbf{x}_{k,1} \right\|_2$; the orientation $\boldsymbol{\lambda}_k$ of the cylinder $V_k$ is defined by the following expression:
$$
\boldsymbol{\lambda}_k = \frac{\mathbf{x}_{k,2}-\mathbf{x}_{k,1}}{l_k}.
$$
Accordingly, the edge $E_k$ of $V_k$ and is given by:
$$
E_k = \left\{\mathbf{x} 
\in \Omega_{\mathrm{v}} \left| \; \mathbf{x}=\mathbf{x}_{k,1}+ s \cdot  \boldsymbol{\lambda}_k,\;s \in \left(0,l_k \right) \right. \right\},
$$
where $s$ denotes the arc length parameter. 
$\Gamma_k$ is the lateral surface of the cylinder $V_k$. The union $\Gamma = \bigcup_{k=1}^N \Gamma_k$ provides an approximation of the surface of the vascular system, due to the fact that the cylinders $V_k$ may not perfectly match to each other. Furthermore, the set $\Lambda = \bigcup_{k=1}^N E_k$ denotes the 1D network composed of the different edges. In the next step, we extract the larger vessels from the microvascular network $\Omega_v$. For this purpose, we choose a fixed threshold $R_T$ and define subsets for the 3D and 1D networks consisting of the larger arterial and venous vessels (see Figure \ref{fig:networks}, right): $\Omega_{lv} = \bigcup_{k=1}^N \left\{ V_k \subset \Omega_{v} \left| R_k>R_T \right. \right\}$ and 
$\Lambda_{lv} = \bigcup_{k=1}^N \left\{ E_k \subset \Lambda \left| R_k>R_T \right. \right\}$. 
The motivation for defining this set is that such vessels may be obtained by means of imaging techniques with acceptable accuracy. Based on $\Omega_{lv}$ and $\Lambda_{lv}$, we construct the surrogate networks.

\subsection{Basic modelling assumptions}
\label{sec:basicmodelling}

Before presenting the modelling equations for blood flow, oxygen transport and vascular growth, 
we discuss the most important modelling assumptions and simplifications, which are fundamental for the choice of the corresponding models. In total, nine assumptions for our modelling concept are formulated:
\begin{itemize}
	\item[(A1)] \textbf{Gravity is neglected.} 
	\ \\ 
	Since the tissue block and the vascular system comprise a volume of about $1.0\;\text{mm}^3$, we can expect that only  little fluid mass is contained in this system. As a result, the gravity force caused by the weight of the blood volume in the upper part of this system has almost no influence on the flow in the lower part.
	\item[(A2)] \textbf{Blood is an incompressible fluid, i.e., its density is assumed to be constant.}
	\ \\ 
	Assumption (A2) does not hold in general, due to the fact that the red blood cells in a microvascular network are not distributed in a homogeneous way \cite{obrist2010red,pries2005microvascular,schmid2015impact}. Besides the red cells, blood is composed of plasma, white cells and platelets \cite[Chap. 1]{formaggia2010cardiovascular}, which implies that on the level of microcirculation blood can not be considered as a homogeneous fluid with a constant density. Despite of this fact, in many publications that are concerned with modelling of blood flow in microvascular networks, blood is considered as an incompressible fluid \cite{cattaneo2014fem,erbertseder2012coupled,peyrounette2018multiscale,possenti2019computational,secomb2013angiogenesis}.
	\item[(A3)] \textbf{The non-Newtonian flow behavior is accounted for by an algebraic relationship.} 
	\ \\ 
	The red blood cells govern the viscosity of blood, significantly. As the red blood cells have to 
	deform such that they can move through capillaries, the viscosity varies within the microvascular network. 
	A quantitative relationship between the vessel diameter $d$ is given by the following formula for the in vivio viscosity 
	$\mu_{\mathrm{bl}}\;\left[ \unit{Pa \cdot s} \right]$ \cite{pries1996biophysical}:
	\begin{equation}
	\label{eq:viscosity}
	\mu_{\mathrm{bl}}\left( d \right) = \mu_{\mathrm{p}} \left( 1+ \left( \mu_{0.45} -1 \right)
	\frac{\left(1-H \right)^C-1}{\left(1-0.45 \right)^C-1} \cdot \left( \frac{d}{d-1.1} \right)^2 \right)
	\cdot \left( \frac{d}{d-1.1} \right)^2.
	\end{equation}
	In \eqref{eq:viscosity}, the diameter $d$ is dimensionless. The physical diameter $\left[\unit{\mu m}\right]$ 
	has to be divided by $1.0\;\unit{\mu m}$ to obtain $d$. $\mu_{\mathrm{p}}\;\left[ \unit{Pa \cdot s} \right]$ denotes the viscosity of blood plasma, and $H$ stands for the discharge hematocrit which is defined by the ratio between the volume of the red blood cells and the total blood volume. The apparent viscosity $\mu_{0.45}$ is given by:
	$$
	\mu_{0.45} = 6.0 \exp\left(-0.085 \cdot d \right) + 3.2 - 2.44 \exp\left(-0.06 \cdot d^{0.645} \right),
	$$
	and $C$ is a coefficient determining the influence of $H$ on $\mu_{\mathrm{bl}}$:
	$$
	C = \left( 0.8 + \exp\left(-0.075 \cdot d \right) \right) \left(-1 + \frac{1}{1 + 10^{-11} d^{12}} \right)
	+ \frac{1}{1 + 10^{-11} d^{12}}.
	$$
	In this context, one should be aware of the fact that the constitutive relationship \eqref{eq:viscosity} is known to hold for human blood. For rat blood, we do not have a suitable function at hand. However, we apply this function also to rat blood. 
	\item[(A4)] \textbf{Inertial effects concerning flows in the vascular system and tissue are not considered.} 
	\ \\ 
	According to \cite[Tab. 1.7]{formaggia2010cardiovascular}, blood velocity is about $0.1\;\unit{mm/s}$ in the arterioles and 
	venules and about $0.01\;\unit{mm/s}$ in the capillary bed in a human system. 
	Therefore, it can be concluded that the Reynolds numbers are significantly lower than $1.0$ and (A4) can be justified \cite{fung1971microcirculation}.
	\item[(A5)] \textbf{Pulsatility of blood flow is neglected.}
	\ \\ 
	The pulsatility of blood flow can be neglected, since the Womersley numbers in the microcirculation are much smaller than $0.1$.
	The Womersley number is given by a dimensionless number comparing the frequency of a pulse and the viscosity of a fluid to each 
	other \cite[Tab. 1.7]{formaggia2010cardiovascular}. 
	\item[(A6)] \textbf{Tissue surrounding the microvascular network is considered as an isotropic, homogeneous porous medium.} 
	\ \\ 
	Considering the tissue, it can be observed that it is mainly composed of cells, fibers and the interstital space  filled with a fluid similar to blood plasma. The interstital space exhibits several pores that are connected by pore throats. Therefore, it is reasonable to consider tissue as a porous medium. In order to model flow within the tissue, we do not use pore network models \cite{weishaupt2019efficient} that attempt to resolve each pore and pore throat, but we apply homogenised or REV-based flow models such as Darcy's law \cite{khaled2003role,possenti2019computational,stoverud2012modeling}. In this context, it is assumed that the  parameters characterising the porous structure like porosity and permeability are constant. Since we do not have any information on the tissue surrounding the given network, we assume that the porosity is constant and that the permeability tensor is diagonal and isotropic with constant values. However, this assumption does not hold for every tissue type. In muscle tissues, for example, the muscle fibers form small channels in which the interstitial fluid  can flow faster than in other space directions. This means that the permeability tensor is not isotropic in this case. If larger domains of an organ like the heart or the liver have to be considered the permeability and porosity of the different tissue layers my vary, due to the fact that different parts of an organ have different functions. A possible way to incorporate this feature into a computational model is to consider a multi-compartment model to simulate flow in tissue composed of multiple layers having a different permeability or porosity. In \cite{hyde2013parameterisation,rohan2018modeling} this modelling approach has been used to simulate vascular systems that are partially homogenised and represented as a porous medium. However, this technique could also be adapted to a heterogeneous tissue.
	
	\item[(A7)] \textbf{No lymphatic drainage is considered.} 
	\ \\ 
    This is due to the fact that we consider a portion of brain tissue. According to medical literature, brain tissue exhibits no lymphatic system as other organs. Over time, several hypotheses have been made on the way interstitial fluid is drained from the brain tissue. Most researchers think that cerebrospinal flow is mainly responsible for the drainage tissue, but the exact exit route of cerebrospinal fluid is still subject of controversies \cite{ahn2019meningeal,lindstrom2018comparison}. Therefore, we do not incorporate lymphatic drainage into the model that is presented in this work. Some numerical models and further information on lymphatic systems in different organs can be found e.g. in \cite{margaris2012modelling,reddy1995mathematical,shvab2017mathematical}.
    	
	\item[(A8)] \textbf{Distribution of oxygen in both the vascular system and tissue is represented	by a continuous unit.}
	\ \\
	In our case, we use the partial pressure of oxygen $PO_2$. The propagation
	of oxygen is governed by diffusion and advection in the vascular system and tissue. 
	In order to model the transport of oxygen in blood, one has to take into account that oxygen molecules are attached to the red blood cells. Therefore, a particle tracer model could be used to determine the oxygen concentration within the vascular network. In this work, we use both for the blood vessel network and the tissue a continuous unit. A commonly used unit to describe the oxygen distribution is the
	partial pressure $\left(PO_2\right)$, which is related to the oxygen concentration by Henry's law \cite{henry1803iii,wagner1973properties}. The consumption of oxygen by the tissue cells is modelled by means of the Michaelis-Menten-law \cite[Chapter 2]{d2007multiscale}\cite{koppl2014influence} containing an average consumption rate and a mean value of the partial pressure in brain tissue.
	
	\item[(A9)] \textbf{New vessels are only added at the boundary nodes of the terminal vessels that are not adjacent to the boundary of the tissue domain.}
	\ \\ 
	Finally, we assume that the creation of new vessels occurs only at the outlets of vessels that are contained in the inner part of the tissue domain. Since our network is formed out of the arterioles and venules which can be detected in the data set, one can imagine that the missing capillary bed can be created by branches or the growing of single vessels at the outlets of the terminal vessels. In medical terms, the formation of new branches at the outlets of blood vessels is referred to as apical growth, while the emergence of new vessels at other places is denoted as vascular growth by sprouting \cite{schneider2012tissue}.
\end{itemize}

\subsection{3D-1D modelling of blood flow in microvascular networks and tissue}
\label{sec:BloodFlow}

Let us consider the network $\Omega_v$. As a first step towards a mathematical model for flow in the vascular network, we decompose $\Omega_v$ into its single vessels $V_k$. Then for each vessel $V_k$, a 1D flow model is derived  (see Step 3 in Figure \ref{fig:DD}). In order to obtain a flow model for the whole network flow, coupling conditions connecting the single vessels as well as boundary conditions are specified  (see Step 4 in Figure \ref{fig:DD}). Finally, a model for tissue flow is determined.

Taking the modelling assumptions (A4) and (A6) from the previous subsection into account, the inertia terms in the standard fluid equations may be dropped. Due to (A5), there is no need to consider complex fluid structure interaction models relating blood flow and deformations of the vessel walls to each other. All in all, it is admissible to simulate both tissue and vascular flow by means of Darcy type equations combined with conservation equations for incompressible fluids. According to \cite{whitaker1986flow}, the permeability $K_{V_k}$ for the vessel $V_k \subset \Omega_v$ is given by: $K_{V_k} = \left( R_k \right)^2/8.$ Using \eqref{eq:viscosity}, the blood viscosity is computed as:
$$
\mu_{\mathrm{bl}}\left( \mathbf{x} \right) = \mu_{\mathrm{bl}}\left( \frac{2  R_k}{1.0\;\mu m} \right)
=: \mu_{\mathrm{bl}}^{(k)} \text{ for } \mathbf{x} \in V_k \subset \Omega_v.
$$
Denoting the pressure in the vascular system by $P_v$, the 3D flow model for a single vessel $V_k$ reads as follows: 
\begin{equation}
\label{eq:DarcyVessel3D}
\mathbf{u}_v = - \frac{K_{V_k}}{\mu_{\mathrm{bl}}^{(k)}} \nabla P_v\;
\text{ in } V_k \subset \Omega_v,
\end{equation}
where $\mathbf{u}_v$ is the velocity field within $V_k$. This relationship is of Darcy or Hagen-Poiseuille type and can be derived from the Navier-Stokes equations assuming a stationary and laminar flow with a parabolic profile (see \cite[Section 2.4]{vcanic2003mathematical} or \cite{mayer1996pressure}).
It remains to specify the boundary conditions for both flow problems. At first, we turn our attention to the boundary conditions on $\Gamma$ coupling flow in the microvascular network and the tissue. 
For this purpose, Starling's filtration law is considered to determine the flux across the vessel surface $\Gamma_k$:
\begin{equation} 
\label{eq:Starling3D}
\mathbf{u}_v \cdot \mathbf{n} = L_p \left( \left( P_v - P_t \right) -\sigma \left( \pi_v-\pi_t\right) \right) =: J_p\left( P_v, P_t \right) \text{ on } \;\Gamma_k,
\end{equation}
where $\mathbf{n}$ is the outward unit vector normal to the vessel surface $\Gamma_k$ and $P_t$ represents the fluid pressure in the tissue. In \eqref{eq:Starling3D}, besides the pressure difference between fluid pressures, the  oncotic pressures $\pi_v$ and $\pi_t$ are also taken into account, where the oncotic pressure difference is weighted by a constant oncotic reflection coefficient $\sigma$. Finally,  both pressure differences are multiplied by the hydraulic conductivity $L_p$ of the vessel wall. For simplicity, we assume that $L_p$ is equal and constant for all the vessels contained in the network $\Omega_v$.

Next, we apply a topological model reduction transforming the 3D flow problem \eqref{eq:DarcyVessel3D} into a 1D flow problem. Since the radius of $V_k,\;k \in \left\{ 1, \ldots,N \right\}$ is small compared to the diameter of $\Omega$, we assume that the function $\mathbf{u}_v$ has a uniform profile with respect to a cross section $\mathcal{D}_k\left(s\right)$ i.e. $P_v$ is constant and equal to $p_v\left(s \right)$ on $\mathcal{D}_k \left(s\right)$ for all $s \in \left(0,l_k \right)$. Integrating over an arbitrary cylinder $Z_k \subset V_k$ having the radius $R_k$ and ranging from $s_1$ to $s_2$, where $s_1<s_2$ and $s_1,\;s_2 \in  \left(0,l_k \right)$, we obtain by \eqref{eq:DarcyVessel3D} and the divergence theorem:
\begin{align*}
0 &= \int_{Z_k}\;\nabla \cdot \left(\frac{K_{V_k}}{\mu_{\mathrm{bl}}^{(k)}} \nabla P_v \right)\;dV = 
\int_{\partial Z_k}\;\frac{K_{V_k}}{\mu_{\mathrm{bl}}^{(k)}} \nabla P_v \cdot \mathbf{n} \;dS = \\
&= -\int_{\mathcal{D}_k \left(s_1\right)} \frac{K_{V_k}}{\mu_{\mathrm{bl}}^{(k)} } \frac{\partial P_v}{\partial s} \;dS 
+ \int_{\mathcal{D}_k \left(s_2\right)} \frac{K_{V_k}}{\mu_{\mathrm{bl}}^{(k)}} \frac{\partial P_v}{\partial s} \;dS
+ \int_{\Gamma_{Z_k}}\;\frac{K_{V_k}}{\mu_{\mathrm{bl}}^{(k)}} \nabla P_v \cdot \mathbf{n} \;dS.
\end{align*}
$\Gamma_{Z_k}$ is the lateral surface of $Z_k$. By means of the interface condition \eqref{eq:Starling3D}, the last surface integral can be further simplified to:
\begin{align*}
-\int_{\Gamma_{Z_k}}\;&\frac{K_{V_k}}{\mu_{\mathrm{bl}}^{(k)}} \nabla P_v \cdot \mathbf{n} \;dS
= \int_{\Gamma_{Z_k}} L_p \left( \left(P_v-P_t\right) -
\sigma \left( \pi_v-\pi_t\right) \right)\;dS = \\
&= L_p \int_{\Gamma_{Z_k}} P_v \;dS - L_p  \int_{\Gamma_{Z_k}} P_t + \sigma \left( \pi_v-\pi_t \right)\;dS = \\
&= 2\;L_p \pi R_k  \int_{s_1}^{s_2} p_v(s)\;ds - L_p \int_{\Gamma_{Z_k}} P_t + \sigma \left( \pi_v-\pi_t \right)\;dS.
\end{align*}
Using the fundamental theorem of integral calculus and $K_{V_k} = R_k^2/8$, we obtain:
\begin{align*}
-\int_{\mathcal{D}_k\left(s_1\right)}\frac{K_{V_k}}{\mu_{\mathrm{bl}}^{(k)} } \frac{\partial P_v}{\partial s} \;dS 
+ \int_{\mathcal{D}_k\left(s_2\right)}\frac{K_{V_k}}{\mu_{\mathrm{bl}}^{(k)}} \frac{\partial P_v}{\partial s} \;dS =
\frac{ R_k^4 \pi}{8 \mu_{\mathrm{bl}}^{(k)} } \int_{s_1}^{s_2} \frac{\partial^2 p_v}{\partial s^2} \;ds.
\end{align*}
Summing up the previous equations, a 1D integral equation for the unknown $p_v$ results: 
\begin{align*}
0 &= \int_{s_1}^{s_2} \frac{R_k^4 \pi}{8 \mu_{\mathrm{bl}}^{(k)} } \frac{\partial^2 p_v}{\partial s^2} 
- 2\;L_p \pi R_k  p_v \;ds + L_p \int_{\Gamma_{Z_k}} P_t + \sigma \left( \pi_v-\pi_t \right)\;dS \\
&=\int_{s_1}^{s_2} \frac{ R_k^4 \pi}{8 \mu_{\mathrm{bl}}^{(k)} } \frac{\partial^2 p_v}{\partial s^2} 
- 2\; L_p \pi R_k p_v + L_p \int_{\partial \mathcal{D}_k\left(s\right)} P_t + \sigma \left( \pi_v-\pi_t \right)\;d\gamma\;ds.
\end{align*}
Since $s_1$ and $s_2$ are chosen arbitrarily, we conclude by the fundamental theorem of variational calculus that the following 1D differential equation holds for each vessel:
\begin{equation}
\label{eq:1DPDE}
0 =\frac{ R_k^4 \pi}{8 \mu_{\mathrm{bl}}^{(k)} } \frac{\partial^2 p_v}{\partial s^2} 
- 2 \pi L_p R_k p_v + L_p \int_{\partial \mathcal{D}_k\left(s\right)} P_t + \sigma \left( \pi_v-\pi_t \right)\;d\gamma. \\
\end{equation}
Using \eqref{eq:Starling3D}, we get:
\begin{equation*}
-\frac{R_k^4 \pi}{8 \mu_{\mathrm{bl}}^{(k)} } \frac{\partial^2 p_v}{\partial s^2} = 
-2 \pi\;R_k J_p\left( p_v, \overline{P}_t \right),
\text{ where }
\overline{P}_t\left(s \right) = \frac{1}{2\pi R_k} \int_{\partial \mathcal{D}_k\left(s\right)} P_t\;d\gamma.
\end{equation*}
This means that the flow equation for $V_k$ is now defined on the 1D centerline $\Lambda_k$ parameterised by 
$s \in \left(0,l_k \right)$. As a consequence, the 3D network $\Omega_v$ is from now on considered as a 1D network $\Lambda$, which is the union of the 1D edges $E_k$. It remains to specify the boundary conditions for $x\in \partial E_k$. Thereby, we consider the following three cases (see Figure \ref{fig:DD}, Step 4):
	\begin{itemize}
	\item \emph{Boundary node:} $x \in  \partial E_k \cap \partial\Lambda$ \\
		At boundary nodes, we set a Dirichlet boundary value: $p_v(x) = p_{boundary}$.
	\item \emph{Inner network node:} $x \in  \partial E_k \wedge x \notin \partial\Lambda$ \\
		At a boundary node located in the inner part of the network $\Lambda$, we enforce the continuity of pressure and the conservation of mass. Let $N\left(x\right)$ be the index set numbering the edges $E_l$ whose boundaries are intersecting  at $x$ i.e.: $x = \partial E_k \cap \partial E_l$. Using this notation, the coupling conditions read as follows:
	\begin{itemize}
			\item Continuity of $p_v$:
			$\;\left. p_v \right|_{x \in \partial E_k} = \left. p_v \right|_{x \in \partial E_l}, \forall l \in N\left(x\right).$ 
			\item Mass conservation:
			$$
			\sum_{l \in N\left(x\right) \cup \left\{k\right\}} \left( \left. -\frac{R_l^4 \pi}{8 \mu_{\mathrm{bl}}^{(l)} } \frac{\partial p_v}{\partial s} \right|_{x} \right)  = 0.
			$$	
	\end{itemize}
	A derivation of these coupling conditions (mass conservation, continuity of pressure) can be found in \cite{formaggia2003one}. In this publication, the coupling conditions for a 1D hyperbolic blood flow problem have been derived by means of an energy inequality. Similar concepts can be applied to derive the coupling conditions for an elliptic flow problem like \eqref{eq:3D1DVascular}.
\end{itemize}
After deriving a simplified flow model for the blood vessel network, we turn our attention to
flow in the porous matrix. Since we reduced our 3D network $\Omega_v$ to a 1D network given by $\Lambda$, we identify the porous tissue matrix $\Omega_t$ with the whole domain $\Omega$. According to (A6), it is assumed that the tissue is an isotropic, homogeneous porous medium. Consequently, the corresponding permeability tensor $\mathbf{K}_t$ is given by a constant scalar $K_t$:
$$
\mathbf{K}_t \left( \mathbf{x} \right) = K_t \cdot \mathbf{I}_3\;\text{ for } \mathbf{x} \in \Omega.
$$ 
Similarly, the viscosity of the fluid in the interstitial space is taken to be constant and is denoted by $\mu_{\mathrm{int}}$. This can be justified by the fact that this fluid consists mostly of water which is a Newtonian fluid. At the boundary $\partial \Omega$, we set for simplicity homogeneous Neumann boundary conditions:
$$
\mathbf{n} \cdot \left( \frac{K_t}{\mu_t} \nabla P_t \right)= 0 \text{ on } \partial \Omega.
$$ 
Posing a no-flow boundary condition implies that the tissue block is supplied by the circulatory system via the arterial blood vessels and drained via the venous part of the network. 

\begin{figure}[h!]
	\begin{center}		
		\includegraphics[width=1.0\textwidth]{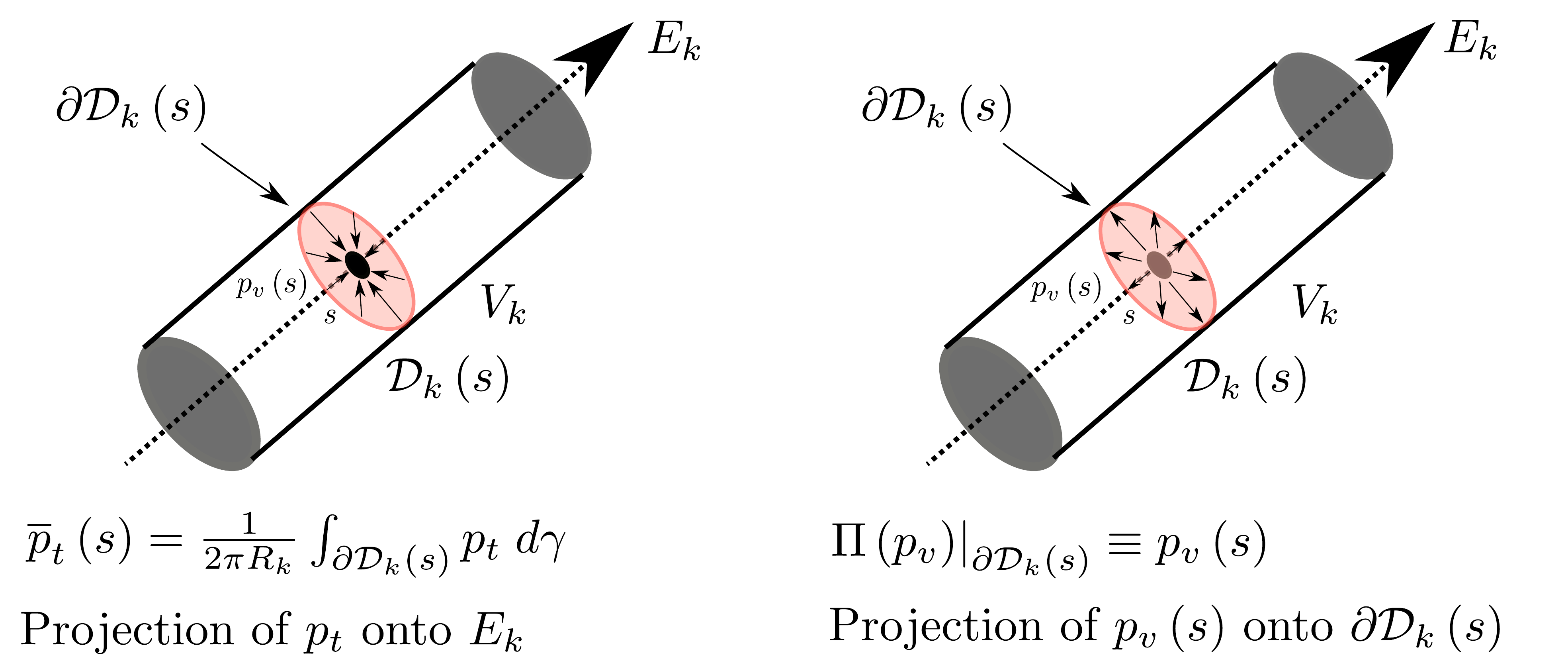}
	\end{center}
	\caption{\label{fig:Projection} In this figure, we illustrate two possible coupling concepts for 3D tissue and 1D vascular flow. On the left hand side, the tissue pressure $p_t$ is averaged and projected on the main axis $E_k$ of the vessel $V_k$. The right hand side shows the projection of the 1D vascular pressure $p_v\left( s \right)$ onto the vessel wall of $V_k$.}
\end{figure}

To model the fluid exchange with the vascular system, we adapt a coupling concept discussed in \cite{cerroni2019mathematical,koppl2018mathematical,kremheller2019approach}. Thereby the influence of the vascular system is incorporated by means of a source term in the corresponding Darcy equation. The source term contains a Dirac measure 
concentrated on the lateral surface $\Gamma$ of the network. This is motivated by the observation that the fluid exchange between the vascular system and tissue occurs across the vessel surface. Another coupling concept consists in concentrating the Dirac source term on the centerline $\Lambda$ of the blood vessel network. However, when using this strategy, line singularities along the centerline \cite{d2007multiscale,d2008coupling} arise, which have to be handled with care. To determine the complete source term for the tissue matrix, we multiply the Dirac measure by Starling's law \eqref{eq:Starling3D}. Combining Darcy's equation and a conservation equation, we have the following flow model for the 3D tissue matrix:
\begin{align}
\label{eq:3DDarcyTissue}
-\nabla \cdot \left( \frac{K_t}{\mu_t} \nabla P_t \right) = J_p\left( \Pi\left(p_v\right), P_t \right) \delta_{\Gamma} \text{ in }\Omega, 
\quad \mathbf{n} \cdot \left( \frac{K_t}{\mu_t} \nabla P_t \right) &= 0 \text{ on } \partial \Omega.
\end{align}
Since $p_v$ is defined on a 1D manifold, we have to project it onto the vessel surface such that the pressure difference between the vascular pressure and the tissue pressure can be computed. For an edge $E_k \subset \Lambda$ the projection operator $\Pi$ occurring in \eqref{eq:3DDarcyTissue} is defined by:
$$
\left. \Pi\left(p_v\right) \right|_{\partial \mathcal{D}_k\left(s\right)} \equiv p_v\left(s \right),\;
\forall \; s \in \left( 0,l_k \right).
$$
Contrary to existing 3D-1D coupled flow models \cite{cattaneo2014fem,cerroni2019mathematical,d2007multiscale,d2008coupling,koppl2018mathematical}, we project by means of the operator $\Pi$ the 1D pressure onto the vessel walls, to compute the pressure differences on $\Gamma_k$. In the aforementioned references, the 3D pressures are projected on the 1D vessels $E_k$, by means of an average operator \cite{d2008coupling} (see Figure \ref{fig:Projection}). 
Summarising \eqref{eq:1DPDE} and \eqref{eq:3DDarcyTissue}, we finally have the following 3D-1D coupled PDE-system governing flow in the vascular system and the tissue:
\begin{align}
\label{eq:3D1DTissue}
- \nabla \cdot \left( \frac{K_t}{\mu_t} \nabla p_t \right) = &\;J_p\left( \Pi\left(p_v\right), p_t \right) \delta_{\Gamma} \text{ in } \Omega, \\
\label{eq:3D1DVascular}
-\frac{ R_k^4 \pi}{8 \mu_{\mathrm{bl}}^{(k)} } \frac{\partial^2 p_v}{\partial s^2} = -2 \pi R_k 
&J_p\left( p_v, \overline{p}_t \right),\;\forall E_k \subset \Lambda,
\end{align}
where 
$J_p\left( P_v, P_t \right) = L_p \left( \left( P_v - P_t \right) -\sigma \left( \pi_v-\pi_t\right) \right).$
Henceforth, we adopt a more uniform notation and denote the 3D tissue pressure as well as the 1D pressure in the vascular system by lower case letters, i.e. $P_t$ is replaced by $p_t$ in \eqref{eq:3D1DTissue}. $p_v$ in \eqref{eq:3D1DVascular} still indicates the 1D blood pressure. A similar 3D-1D coupled flow model can be found in \cite{kojic2017composite}. One major difference between the PDE-system in \eqref{eq:3D1DTissue} and \eqref{eq:3D1DVascular} and the model in \cite{kojic2017composite} is the type of coupling vascular and tissue flow. Our model is based on Dirac source terms in 3D which are concentrated on the vessel surfaces. In \cite{kojic2017composite} this is not the case. Here, the 3D source term is concentrated on the center lines of the microvascular network. However, modelling the exchange processes between tissue and vascular system based on the vessel surfaces as it is done in \eqref{eq:3D1DTissue} and \eqref{eq:3D1DVascular} is more realistic, since the exchange processes occur at the permeable vessel surfaces.
		
\subsection{3D-1D modelling of oxygen transport in microvascular networks and tissue}
\label{sec:transport}

To determine the oxygen distribution in both the blood vessel network and the surrounding tissue matrix, we assign to each system a variable quantifying the partial pressures. In the following, we denote them by $P_{O_2}^{(v)}$ for the vascular system and by $P_{O_2}^{(t)}$ for the tissue block. As the fluid pressures, $P_{O_2}^{(v)}$ is defined on the 1D network $\Lambda$, and $P_{O_2}^{(t)}$ is defined on $\Omega$. Concerning the distribution of the partial pressures, we assume that it is governed by convection and diffusion, where the diffusion coefficients for the vascular system and the tissue are given by two constants $D_v$ and $D_t$.

As in Section \ref{sec:BloodFlow}, we use a 3D-1D coupled model for these processes. The derivation is similar to the one developed for the blood flow problem. For convenience, we briefly describe here the most important steps: First, the network is decomposed and a 1D PDE for each vessel is derived. Next, we assume that the vascular partial pressure is assumed to be constant across the section area. Based on this assumption a 3D convection-diffusion equation for each vessel is reduced to a 1D convection-diffusion equation using similar arguments as in Subsection \ref{sec:BloodFlow}. After that suitable coupling and boundary conditions are chosen for the boundary nodes of the 1D network.

The 3D model for the tissue is again defined on $\Omega$. Similarly for the blood flow process, the exchange between tissue and vascular system is accounted for by means of a Dirac measure concentrated on the lateral surfaces $\Gamma_k \subset \Gamma$. Modelling the flux of oxygen from the vessels into the tissue matrix, we consider the vessel walls as semipermeable membranes allowing a selective filtration of molecules. According to \cite[Section 5.2]{cattaneo2014fem} the Kedem-Katchalsky equation is a suitable model to determine the flux $J_{P_{O_2}}$ of oxygen across the vessel wall:
\begin{align}
\label{eq:KedemKatchalsky}
J_{P_{O_2}} \left( \Pi\left(p_v\right),p_t,\Pi\left( P_{O_2}^{(v)} \right), P_{O_2}^{(t)} \right) &= \left(1-\sigma \right) 
J_p\left( \Pi\left(p_v\right), p_t \right) P_{O_2}^{(t/v)} \\
\nonumber
&+  L_{P_{O_2}} \left( \Pi\left( P_{O_2}^{(v)} \right) - P_{O_2}^{(t)} \right), \;\forall\; \Gamma_k \subset \Gamma.
\end{align}
The symbol $P_{O_2}^{(t/v)}$ denotes the average partial pressure in vessel walls. In this work, we define it as the arithmetic mean of the partial pressures on the two sides of the walls \cite[Chapter 5.2]{cattaneo2014fem} \cite{perktold2009mathematical}:
$$
P_{O_2}^{(t/v)} = \frac 12 \left( P_{O_2}^{(t)} + \Pi\left( P_{O_2}^{(v)} \right) \right),
$$
The parameter $L_{P_{O_2}}$ in \eqref{eq:KedemKatchalsky} indicates the permeability 
of the vessel wall with respect to oxygen. As mentioned in Section \ref{sec:basicmodelling}, we use the Michaelis-Menten formula \cite{secomb2013angiogenesis},
\begin{equation}
\label{eq:MichaelisMenten}
m^{(ox)}\left( P_{O_2}^{(t)} \right)  = \frac{m_0^{(ox)}}{P_{O_2}^{(t)}+P_{O_2,0}^{(t)}}P_{O_2}^{(t)}
\end{equation}
to model the oxygen consumption of the tissue cells. $m_0^{(ox)}$ represents the maximal oxygen demand. For this work, we assume that it is 
constant. $P_{O_2,0}^{(t)}$ is the partial pressure at half maximal consumption. In this work, we assume that both parameters are constant. Summarising the considerations of this subsection and using \eqref{eq:KedemKatchalsky} and \eqref{eq:MichaelisMenten}, we obtain the following 3D-1D coupled model for oxygen transport:
\begin{align}
\label{eq:TissuePO23D1D}
\nabla \cdot \left( \mathbf{u}_t  P_{O_2}^{(t)} - D_t \nabla P_{O_2}^{(t)} \right) = 
-m^{(ox)}\left( P_{O_2}^{(t)} \right) \hspace{4.5cm}& \\
+ J_{P_{O_2}}\left(\Pi\left(p_v\right),p_t,\Pi\left( P_{O_2}^{(v)} \right), 
\nonumber
P_{O_2}^{(t)} \right)\delta_{\Gamma} \text{ in }& \Omega, \\
\label{eq:VascularPO23D1D}
\frac{\partial}{\partial s}\left( u_v P_{O_2}^{(v)} - D_v \frac{\partial P_{O_2}^{(v)}}{\partial s}  \right) = -2\pi R_k
J_{P_{O_2}}\left(p_v,\overline{p}_t,P_{O_2}^{(v)},\overline{P_{O_2}^{(t)}}\right), \;\forall E_k \subset \Lambda,\\
\nonumber
\mathbf{n} \cdot \left( \mathbf{u}_t  P_{O_2}^{(t)} - D_t \nabla P_{O_2}^{(t)} \right) = 0\text{ on } \partial\Omega.
\end{align}
The velocities $u_v$ and $\mathbf{u}_t$ are computed by means of Darcy's law:
\begin{equation}
\label{eq:DarcyVelocities}
\left. u_v \right|_{\Lambda_k} = -\frac{R_k^2}{8 \mu_{\mathrm{bl}}^{(k)} } \frac{\partial p_v}{\partial s}\; 
\text{ and } \;\mathbf{u}_t = -\frac{K_t}{\mu_t} \nabla p_t.
\end{equation}
As for the fluid problem, we and use Dirichlet conditions for the boundary nodes $x \in  \partial E_k \cap \partial\Lambda$. Thereby, we have to check first, if $E_k$ is part of a vein or an artery (see also Subsection \ref{sec:Simulation_parameters}). Depending on the decision a typical value for a venous or an arterial partial pressure is chosen. In this work, we compute the fluid pressures and velocities and consider a vessel as vein, if its fluid pressure or velocity is below the corresponding average. This is motivated by the fact that veins belong to the low pressure and velocity region in the vascular system. In the other case, the vessel is considered as an artery. At the inner networks nodes $\left( x \in  \partial E_k \wedge x \notin \partial\Lambda \right)$, the continuity of partial pressures and a mass conservation equation is considered. In order to formulate the coupling equations, we use the same notation as for the fluid problem and the coupling conditions for the partial pressure of oxygen at inner network nodes read as follows:
\begin{itemize}
	\item Continuity of $P_{O_2}^{(v)}$:
	$\left. P_{O_2}^{(v)} \right|_{x \in \partial E_k} = \left. P_{O_2}^{(v)} \right|_{x \in \partial E_l}, \forall l \in N\left(x\right),$
	\item Mass conservation:
	$$
	\sum_{l \in N\left(x\right) \cup \left\{k\right\}} \left. \left( u_v P_{O_2}^{(v)} - D_v \frac{\partial P_{O_2}^{(v)}}{\partial s} \right) \right|_{x \in \partial E_l}  = 0.
	$$
\end{itemize}
\newpage
\subsection{Generation of surrogate microvascular networks}
\label{sec:GenerationNetworks}

The process of generating a surrogate microvascular network out of the well-segmented arterioles and venoles is divided into three different phases:
\begin{itemize}
	\item[]\textbf{Phase 1 (P1)} Generation of the larger blood vessels i.e. blood vessels with a radius ranging from about $4.5\;\mu m$ to $16.0\;\mu m$.
	\item[]\textbf{Phase 2 (P2)} Generation of the fine scaled vessels i.e. blood vessels with a radius
	ranging from about $2.0\;\mu m$ to $4.5\;\mu m$.
	\item[]\textbf{Phase 3 (P3)} Removal of terminal vessels ending in the inner part of the tissue block.
\end{itemize}
This is motivated as follows: Since even a microvascular network like the one depicted in Figure \ref{fig:networks}, exhibits a multiscale structure, it is reasonable to first generate the large scaled vessels in a first step. Otherwise, the growth of the larger vessels might be blocked by some smaller vessels and the main vessels of the network can not be developed. In a second step, we add the fine scaled vessels sprouting out of the large scaled vessels. 
\begin{algorithm}[h!]
	\caption{\label{alg:collision} Collision detection algorithm}	
	\SetAlgoLined
	\ \\
	\textbf{Preliminary work:} The computational domain $\Omega$ is decomposed into eight disjunctive sub-control volumes in form of cuboids $SV_l,\;l\in \left\{1,\ldots,8\right\}$. Next the edges of the 1D network $\Lambda$ are assigned to  the $SV_l$ in which they are contained. If a new edge $E_{new}$ is contained in $SV_l$ only the edges in $SV_l$ have to be checked for intersection and a loop over all edges in $\Lambda$ is avoided\;
	\textbf{INPUT:} New edge $E_{new}$ to be added to $\Lambda$, 1D network $\Lambda$, sub-control volumes $SV_l,\;l \in \left\{1,\ldots,8\right\}$ \;
	Determine the sub-control volumes $SV_l$ with $E_{new} \cap SV_l \neq \emptyset$ \;
	Add $E_{new}$ to the detected sub-control volumes $SV_l$ \;
	Get radius $R_{new}$ of the edge $E_{new}$ \; 
	isIntersecting = \textbf{false} \;
	\For{all edges $E_k \cap SV_l \neq \emptyset$}{
		Get radius $R_k$ of the edge $E_k$ \;
		Compute the eucledian distance $\text{dist}_{nk}$ between $E_{new}$ and $E_k$\;
		\If{$E_{new}$ and $E_k$ do not share a common network node AND $\text{dist}_{nk}<R_{new}+R_k$}{
			\textbf{OUTPUT:} $E_{new}$ is intersecting an existing edge\;
			isIntersecting = \textbf{true} \;
			Remove $E_{new}$ from $SV_l$ \;
			\textbf{BREAK} \;
		}
	}
	\If{!isIntersecting}{
		\textbf{OUTPUT:} $E_{new}$ is intersecting no existing edge\;
		Add $E_{new}$ to $\Lambda$ \;
	}
\end{algorithm}	
The lower bound of $2.0\;\mu m$ for the vessel radii in \textbf{(P2)} is motivated by considering the distribution of radii in Figure \ref{fig:segmented}. It can be observed that most of the radii are larger than $2.0\;\mu m$. Therefore, we choose this value as a lower bound in \textbf{(P2)}. In order to prevent an excessive runtime of our iteration procedure, we restrict the total number of growth steps for (P1) and (P2) by $N_{it,\max}^{(1,2)}=35$ and for (P3) by $N_{it,\max}^{(3)}=15$. Both the generation of the larger vessels and the generation of the smaller vessels is carried out in an stepwise manner (see Subsection \ref{sec:basics}), where new vessels are only added at the outlets of the terminal vessels. Thereby, both arterial and venous vessels are updated.

However, after finishing Phase \textbf{(P2)}, it can be observed that in the inner part of the tissue there are several vessels, which are not fully integrated in the network. In a healthy microvascular network this is usually not the case, since it exhibits fine scaled vessels or a capillary bed forming continuous connections between the arterial and venous part of the microvascular network. An important tool that is used several times in the algorithm below, is a method to detect colliding or intersecting vessels (see Algorithm \ref{alg:collision}). Avoiding the intersection of blood vessels is crucial for generating a well structured blood vessel network. In the remainder of this subsection, we provide the key ingredients for the three phases listed above:
\begin{itemize}
	\item \textbf{Phase 1 (P1):}
	\ \\ 
	Given a 1D network $\Lambda_j$ from the $j$-th step in (P1), we compute the fluid pressures in $\Omega$ as described in Subsection \ref{sec:BloodFlow}. Please note that for $j=0$, $\Lambda_j$ is given by the large scaled vessels $\Lambda_{lv}$ presented in Figure \ref{fig:networks} (right). Based on the velocity fields, the distribution of the partial pressures in both tissue and vascular system is determined using the PDEs provided in Subsection \ref{sec:transport}. Next, we add new blood vessels at the boundary nodes of the network $\Lambda_j$. This results in a new 1D vascular network $\Lambda_{j+1}$ for the next step. Let us consider a boundary node $x$ contained in an edge $E_k^{(j)} \subset \Lambda_j$: $x \in \partial \Lambda_j \cap \partial  E_k^{(j)}$. For each boundary node $x$ the normalised gradient of $P_{O_2}^{(t)}$ is determined: 
	\begin{equation}
	\label{eq:PO2Direction}
	d_p\left(x\right) = \frac{\nabla P_{O_2}^{(t)}\left(x\right)}{\left\| \nabla P_{O_2}^{(t)} \left(x\right)\right\|_2}.	
	\end{equation}
	Choosing $d_p$ as the new direction of growth reflects the property of microvascular networks to grow towards the tissue region with an increased metabolic demand. However, our numerical tests showed that following strictly $d_p$ can result into sharp bendings posing a high resistance on 
	blood flow. Therefore, we follow an idea presented in \cite{schneider2012tissue}. In this publication, the orientation $d_k$ of $E_k^{(j)}$ is multiplied with a regularisation parameter $\lambda_g$ and combined with $d_p$ to avoid such sharp bendings:
	\begin{equation}
	\label{eq:bendings}
	d_g\left(x\right) = d_p\left(x\right) + \lambda_g \cdot d_k\left(x\right), 
	\end{equation}	
	where $d_g\left(x\right)$ denotes the new direction of growth at the node $x$. Normalising the vector $d_g\left(x\right)$, we obtain the new growth direction. It remains to specify the radius and length of the new segment. If a single vessel is generated out of $E_k^{(j)}$, the vessel radius is taken over from the father segment $E_k^{(j)}$. To determine the length $l$ of the new vessel, we assume a log-normal distribution of the ratio $r = l/R$ between the vessel length and the vessel radius: 
	\begin{equation}
	\label{eq:log_ratio}
	\log \mathcal{N}\left(r,\mu_r,\sigma_r\right) = 
	\frac{1}{r \sqrt{2\pi \sigma_r^2}} \exp\left( -\frac{\left( \ln r - \mu_r\right)^2}{2\sigma_r^2}\right),
	\end{equation}
	where $\mu_r$ and $\sigma_r$ denote the mean value and standard deviation of the log-normal distribution, respectively. In accordance with the distribution in \eqref{eq:log_ratio}, we compute $r$. Using the radius $R_k$ of the father segment $E_k^{(j)}$, the length $l$ of the new vessel is given by: $l = R_k \cdot r$.

	Besides the formation of a single vessel at the inner boundary of $E_k^{(j)}$, we consider also the possibility of generating a bifurcation. As in \cite{schneider2012tissue}, the cumulative distribution function $P_{bif}$ of \eqref{eq:log_ratio} is used to decide whether a bifurcation or a single vessel is created. The function $P_{bif}$ is given by the following expression:
	\begin{equation}
	\label{eq:Pbif}
	P_{bif}\left(r\right) = \Phi \left( \frac{\ln r - \mu_r}{\sigma_r} \right) = \frac 12 + \frac 12 \text{erf}\left( \frac{\ln r-\mu_r}{\sqrt{2\sigma_r^2}}\right).
	\end{equation}
	In this context, $\Phi$ represents the standard normal cumulative distribution function and $\text{erf}$ the Gaussian error function. If $P_{bif}\left(r\right)$ exceeds for $E_k^{(j)}$ a certain threshold $P_{th}=0.6$, a bifurcation is attached to the boundary node. We observed that this threshold leads to a reasonable number of bifurcations in the surrogate network. If $P_{bif}\left(r\right) \leq P_{th}$ a single vessel is generated as described before. For our simulations, we used $\mu_r = 2.4$ and $\sigma_r = 0.3$. The choice of $\mu_r$ and $\sigma_r$ results from several numerical simulations. Using these values our vascular trees exhibit radii and lengths in a reasonable range, as we will discuss in Section \ref{sec:SimulationResults}. At this point, we deviate from \cite{schneider2012tissue}. The authors of this publication use the following values: $\mu_r = 14.0$ and $\sigma_r =  6.0$. Now the issue arises of, how to choose the radii, orientations and lengths of the new branches $b_1$ and $b_2$. The radii of the new branches are computed based on a Murray type law, which relates the radius $R_k$ of the father vessel to the radius $R_{b_1}$ of branch $b_1$ and the radius $R_{b_2}$ of branch $b_2$ \cite{murray1926physiological}:
	\begin{equation}
	\label{eq:Murray}
	R_k^{\gamma} = R_{b_1}^{\gamma} + R_{b_2}^{\gamma},
	\end{equation}
	where $\gamma$ denotes the bifurcation exponent. However, it is not quite clear, how to choose this parameter. In \cite{fung1997biomechanics,murray1926physiological,schneider2012tissue,zamir2002physics} one can find values for $\gamma$ ranging from $2.0$ to $3.5$. Our numerical simulations show that this parameter influences the shape of the microvascular network significantly. Therefore, we study in Subsection \ref{sec:results} how a variation of this parameter affects the morphology of the network. Besides \eqref{eq:Murray}, a further equation is required to determine the radii of the branches. According to \cite{schneider2012tissue}, we assume that $R_{b_1}$ follows a Gaussian normal distribution:
	\begin{equation}
	\label{eq:radiusbranch1}
	R_c = 2^{ -\frac{1}{\gamma} }R_k,\;R_{b_i}  \sim \mathcal{N}\left( R, \mu = R_c, \sigma = R_c/32 \right),\;i \in \left\{1,2\right\}.
	\end{equation}
	The heuristic choice of $R_{b_i}$ has the following interpretation: Based on the radius $R_k$ of the father vessel, we compute the expected radius $R_c$ resulting from Murray's law for a fully symmetric bifurcation $\left( R_{b_1}=R_{b_2} \right)$. $R_c$ is used as a mean value for a Gaussian normal distribution, with a small standard derivation. By this, we obtain bifurcations that are slightly deviating from a symmetric bifurcation whose radii are chosen in accordance with Murray's law.
	Having $R_{b_1}$ and $R_{b_2}$ at hand, we compute the corresponding lengths $l_{b_1}$ and $l_{b_2}$ as in the case of a single vessel. The creation of a bifurcation is accomplished by specifying the orientations of the two branches. At first, we define the plane, in which the bifurcation is contained. The normal vector $\mathbf{n}_p$ of this plane is given by the cross product of the vessel orientation $d_k$ and the growth direction $d_g$ of the non-bifurcating case:
	$$
	\mathbf{n}_p = \frac{d_k \times d_g}{\left\| d_k \times d_g \right\|_2}.
	$$
	The exact location of the plane is determined such that the vessel $E_k^{(j)}$ is contained in this plane. Further constraints for the bifurcation configuration are related to the bifurcation angles. In
	\cite{murray1926physiological2} and \cite{rosen2013optimality}, it is shown how optimality principles of minimum work and minimum energy dissipation can be used to derive formulas relating the radii of the branches to the branching angles $\phi_k^{(1)}$ and $\phi_k^{(2)}$:
	\begin{equation}
	\label{eq:bifurcation_angles}
	\cos\left( \phi_k^{(1)} \right) = \frac{R_k^4+R_{b_1}^4-R_{b_2}^4}{2 \cdot R_k^2 R_{b_1}^2} \;\text{ and }\;
	\cos\left( \phi_k^{(2)} \right) = \frac{R_k^4+R_{b_2}^4-R_{b_1}^4}{2 \cdot R_k^2 R_{b_2}^2}.
	\end{equation}
	$\phi_k^{(i)}$ denotes the bifurcation angle between branch $i,\;i \in \left\{1,2\right\}$ and the father vessel (see Figure \ref{fig:creatingbifurcation}, right). Rotating the vector $d_k$ at $x$ around the axis defined by $\mathbf{n}_p$ counterclockwise by $\phi_k^{(1)}$ and clockwise by $\phi_k^{(2)}$, we obtain two new growth directions $d_{b_1}$ and $d_{b_2}$. These vectors can be used to define the main axes of the two cylinders representing the two branches. However, this approach does not allow us to incorporate the direction of the local oxygen gradient into the growth direction. For this reason the optimality conditions \eqref{eq:bifurcation_angles} for the bifurcation angles are relaxed. This is done by choosing the vector $d_{b_i},\;i \in \left\{1,2\right\}$ minimising the difference to $d_g$:
	$d_{\min}^{(k)} = \text{argmin}_{d_{b_i}} \left\|  d_{b_i} - d_g \right\|_2.$
	As orientations for the bifurcation, we keep the direction $d_{b_{j_1}} \neq d_{\min}^{(k)},\;j_1 \in \left\{1,2\right\}$, while the other orientation is given by a vector $d_{b_{j_2}}$ splitting the angle between $d_{\min}^{(k)}$ and $d_g$ (see Figure, \ref{fig:creatingbifurcation}, right) in two halves. 
	This choice for the second growth direction can be considered as a compromise between the optimality principles provided by \eqref{eq:bifurcation_angles} and the tendency of the network to adapt its growth direction to the oxygen demand of the surrounding tissue. Before the new vessels are attached to the inner boundaries of $\Lambda_j$, Algorithm \ref{alg:collision} is used to test whether the newly created vessels are intersecting an existing vessel.
	If a vessel is attached to a boundary node $x$, we obtain a new boundary node $y$ for $\Lambda_{j+1}$. The new boundary values for $p_v$ and $PO_2^{(v)}$ are taken over from the old boundary node $x$. In the final step of \textbf{(P1)}, we check, whether there is a terminal vessel left at which a large scaled vessel can be attached. A terminal vessel $E_k^{(j+1)} \subset \Lambda_{j+1}$ is considered as a large scaled vessel, if its radius $R_k$ is larger than $4.5\;\mu m$. 
	In addition, it is observed whether the $PO_2^{(t)}$ distribution is changing significantly. In order to test this, we consider the region of interest $\Omega_{roi} \subset \Omega$ and compute an averaged value $PO_{2,roi}^{(t)}$:
	\begin{equation}
	\label{eq:PO2_roi}
	PO_{2,roi}^{(t)} = \frac{1}{\left| \Omega_{roi} \right|} \int_{\Omega_{roi}} PO_2^{(t)}\;dV.
	\end{equation}
	If the relative change of $PO_{2,roi}^{(t)}$ with respect to the previous iteration is below $1.0\;\%$, we consider the $PO_2^{(t)}$ distribution as stationary. Phase (P1) is summarised in Algorithm \label{alg:phase1} and the result of (P1) is denoted by $\Lambda_{P1}$.
	\begin{figure}[h]
	\begin{center}		
	\includegraphics[width=1.0\textwidth]{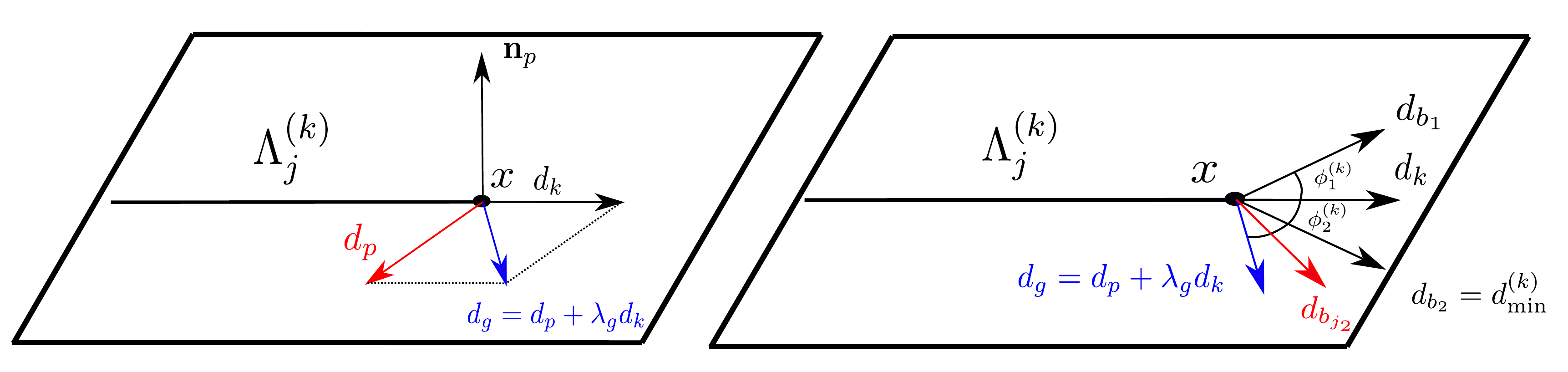}
	\end{center}
	\caption{\label{fig:creatingbifurcation} The picture on the left hand side shows the plane defined by the normal $\mathbf{n}_p$ and the boundary node $x$. Furthermore the orientation vector $d_k$ of the vessel $E_k^{(j)}$ is shown. $d_p$ represents a vector pointing into the direction of the local gradient of the partial pressure at $x$. It can be seen that $d_p$ forms a sharp bending at $x$ with respect to $E_k^{(j)}$. A linear combination of $d_p$ with $\lambda_g d_k$ results in a smoothing of this sharp corner. Thereby, $\lambda_g$ plays a role of a regularisation parameter. On the right hand side the two bifurcation angles $\phi_k^{(1)}$ 
	and $\phi_k^{(2)}$ resulting from Murray's law \eqref{eq:bifurcation_angles} are shown. The vector $d_{b_{j_2}}$ represents the direction of growth resulting from the regularised vector $d_g$ and the direction $d_{b_2}$ obtained from Murray's law. It can be considered as a compromise between the metabolic demand and the optimality principles given by \eqref{eq:bifurcation_angles}.}
	\end{figure}
	\begin{algorithm}
		\caption{\label{alg:phase1} Algorithm for Phase 1}	
		\SetAlgoLined
		\textbf{INPUT:} 1D network $\Lambda_{lv}$ consisting of the larger arteries and veins, domain $\Omega$ \;
		Set $j = 0$, $\Lambda_j \leftarrow \Lambda_{lv}$, $PO_{2,roi,old}^{(t)} = 0$,
		StopPhase1 = \textbf{false} \;
		\While{!StopPhase1}{
			Determine the pressures $p_v$ and $p_t$ by solving \eqref{eq:3D1DTissue} in $\Omega$ and \eqref{eq:3D1DVascular} in $\Lambda_j$\;
			Compute the partial pressures $P_{O_2}^{(v)}$ and $P_{O_2}^{(t)}$  by solving \eqref{eq:3D1DTissue} in $\Omega$ and \eqref{eq:3D1DVascular} in $\Lambda_j$\;
			\For{all boundary nodes $x \in \Omega \cap \partial \Lambda_j $}{
				Let $x \in \partial E_k^{(j)}$, get radius $R_k$ of edge $E_k^{(j)} \subset \Lambda_j$ \;
				\If{$R_k>4.5\;\mu m$}{
					Compute the growth direction $d_g = d_p + \lambda_g d_k$ according to \eqref{eq:bendings}\;
					Compute the length and radius of the new vessel according to \eqref{eq:log_ratio}\;
					Evaluate $P_{bif}$ according to \eqref{eq:Pbif}\;
					\eIf{$P_{bif}>P_{th}$}{
						Create a second branch by means of Murray's laws \eqref{eq:Murray}, \eqref{eq:radiusbranch1} and \eqref{eq:bifurcation_angles}\;	
						Intersection tests (Alg. \ref{alg:collision}) for the new vessels, adapt the boundary conditions\;			
					}{
						\vspace{-0.2cm}
						Intersection test (Alg. \ref{alg:collision}) for the new vessel, adapt the boundary conditions\;			
					}
				}
			}

			Add the new vessels and update the network $\Lambda_{j+1} \leftarrow \Lambda_j$, compute $PO_{2,roi,j}^{(t)}$ according to \eqref{eq:PO2_roi}, determine the number of the large terminal vessels $\left(N_{term,lv} \right)$ \;
			\If{$\left|PO_{2,roi,j}^{(t)}-PO_{2,roi,old}^{(t)}\right|/PO_{2,roi,j}^{(t)}<1.0 \cdot 10^{-2}$ OR $j>N_{it,\max}^{(1)}$ OR $N_{term,lv} = 0$}{
				StopPhase1 = \textbf{true}\;
			}

			$PO_{2,roi,old}^{(t)} = PO_{2,roi,j}^{(t)}$, $\;j = j+1$\;
		}

		\textbf{OUTPUT:} Network containing the main vessels: $\Lambda_{P1} \leftarrow \Lambda_j$\;
	\end{algorithm} \ \\
	
	\item \textbf{Phase (P2):} 
	\ \\
	Having the 1D network $\Lambda_{P1}$ from the first growth phase at hand, we compute as in (P1) the fluid pressures as described in Subsection \ref{sec:BloodFlow}. Based on the resulting velocity fields, the distribution of the partial pressures in both tissue and vascular system is determined using the PDEs provided in Subsection \ref{sec:transport}. The third step of the second phase uses essentially the same concepts presented in (P1). An exception is the choice of the radii. If the computed radius is below a threshold of $3.0\;\mu m$, we deviate from the strategy discussed in (P1). This is motivated by the observation that a huge number of vessels becomes too tiny, if we follow strictly the steps of (P1). Instead, it is assumed that a new radius $R_k<3.0\;\mu m$ follows another normal distribution: 
		\begin{equation}
		\label{eq:SmallRadii}
		R_k \sim \mathcal{N}\left( R, \mu = 2.75\;\mu m, \sigma = 0.25\;\mu m \right).
		\end{equation} 
		By this choice, the distribution of the radii is mainly concentrated on the range of the fine scale vessels i.e. $\left[1.5\;\mu m,4.5\;\mu m\right]$.	In addition to that, we demand that $2.0\;\mu m \leq R_k \leq R_p,$ where $R_p$ denotes the radius of the father vessel. As it has already been mentioned, this is motivated by the fact that almost all the vessels of the segmented network are larger than $2.0\;\mu m$ (see Figure \ref{fig:segmented}). The upper bound $R_k \leq R_p$ ensures that the radius $R_k$ of the new vessel does not exceed the radius $R_p$ of the father vessel. Applying this strategy, the main part of the vessel radii are within a reasonable range and the distribution of the radii has a smooth shape. Before the new vessels are attached to the boundaries of $\Lambda_j$, we test as in (P1.4) if a new vessel intersects an existing vessel or the boundary $\partial \Omega$. Depending on the result it is decided whether a certain vessel can be integrated into the new network. By this, we obtain a new network $\tilde{\Lambda}_j$. If a vessel is attached to a boundary node $x$, we obtain a new boundary node $y$ for $\tilde{\Lambda}_j$. The new boundary values for the blood pressure $p_v$ and the partial pressure of oxygen $PO_2^{(v)}$ are taken over from the old boundary node $x$.
		\begin{figure}[h!]
			\begin{center}		
				\includegraphics[width=0.8\textwidth]{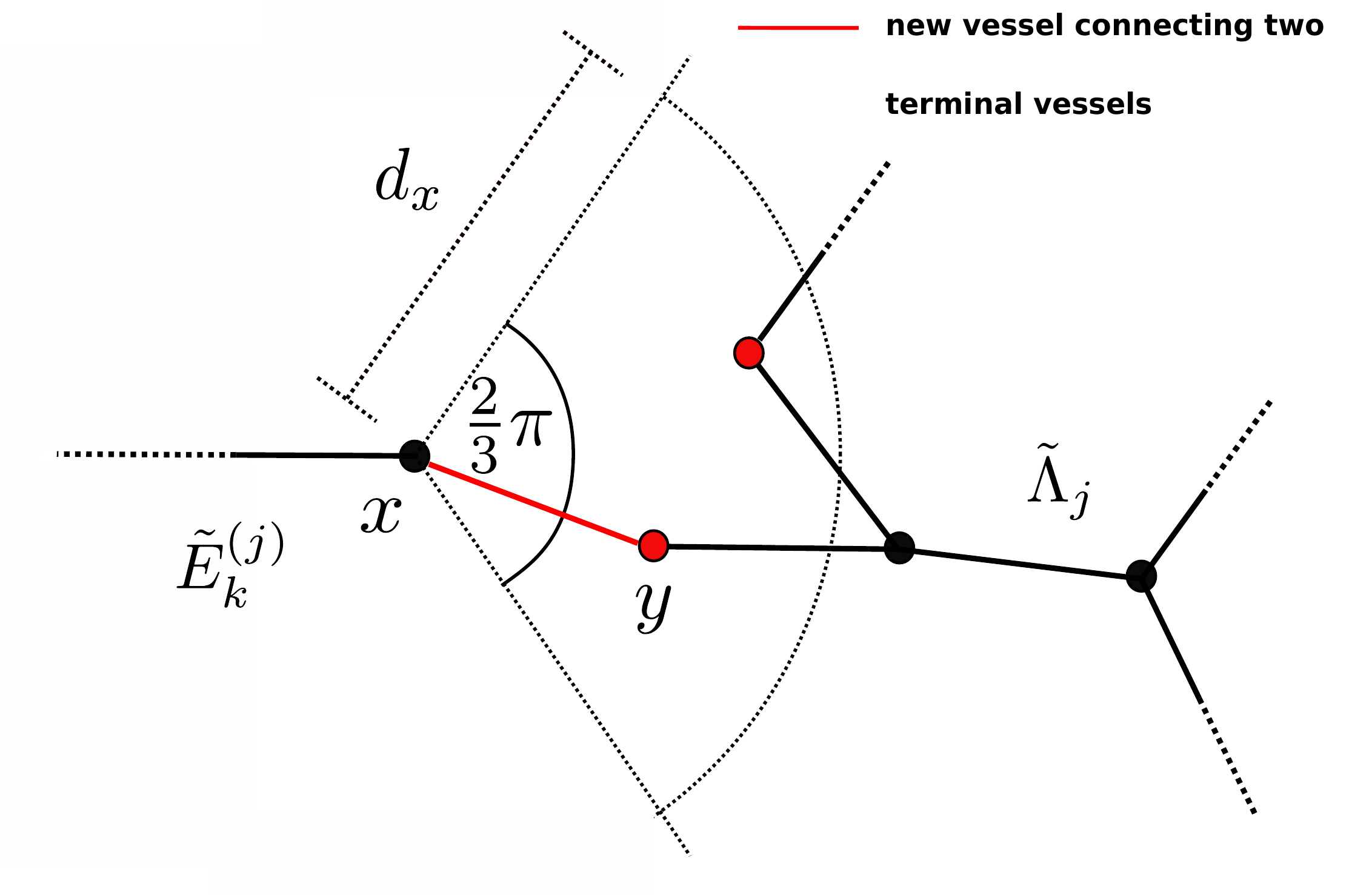}
			\end{center}
			\caption{\label{fig:Connection} This figure shows a terminal vessel $\tilde{E}_k^{(j)}$ with a boundary node $x$. For convenience, we show a 2D sketch of a part of the network $\tilde{\Lambda}_j$. Connecting the boundary node $x$ with other network nodes that are contained in $\tilde{\Lambda}_j$, we consider only those network nodes that are within a cone of opening angle $\frac{2}{3}\pi$. From these network nodes, we choose those that have a distance less than $d_x$ from $x$ (red nodes) and connect one of them with $x$. Thereby, we consider preferably nodes of minimal distance to $x$ and of maximal pressure difference between $x$ and the other node. The node connected with $x$ is denoted by $y$, such that $x$ and $y$ form a new blood vessel that can be added to the network $\Lambda_{j+1}$ (red segment).}
		\end{figure}
		In a next step, we check if the new inner boundaries of $\tilde{\Lambda}_j$ can be connected with another network node. For this purpose the network nodes in the vicinity of a boundary node $x \in \partial \tilde{\Lambda}_j$ are taken into account. Thereby, only network nodes that are contained in a cone with an opening angle of $2 \pi/3$ are considered \cite{secomb2013angiogenesis}. The axis of the cone is given by the orientation of the terminal vessel containing $x$. By this convention, the formation of sharp bendings can be circumvented. From the network nodes within the cone, we choose only those network nodes as candidates that have at most a distance $d_x$ to $x$. In order to create a smooth distribution of the vessel lengths, it is assumed that $d_x$ can be chosen according to the following distribution:
		$$
		d_x \sim \mathcal{N}\left( d_x, \mu = 60.0\;\mu m, \sigma = 10.0\;\mu m \right).
		$$
		This is motivated by the fact that blood vessels in a microvascular network grow at least about $50.0\;\mu m$ per day \cite{pries2014making}. Among the network nodes fullfilling these conditions, we search for those with minimal distance and the highest pressure difference. The pressure difference is computed by means of the pressure value at $x$ and the pressure value at the network node that could be connected with $x$. As a consequence, we link preferably vessels from the venous and the arterial side of the microvascular network, in order to enable a circulation of blood within the microvascular system. The radius of the new segment or blood vessel is given as the arithmetic mean of the terminal vessel containing $x$ and the segment containing the other network node (see Figure \ref{fig:Connection}). Next, we check by means of Algorithm \ref{alg:collision}, whether they intersect other vessels and adapt the boundary conditions. In the final step of (P2), $\Omega_{roi}$ is decomposed in $N_{cv}$ control volumes $CV_i$ having the shape of a cuboid: $\Omega_{roi} = \bigcup_{i=1}^{N_{cv}}CV_i$. For our computations, we used in each space direction $4$ control volumes i.e. $N_{cv}=64$. With respect to each $CV_i$ and $\Omega_{roi}$, we compute averaged $PO_2^{(t)}$ values:
		\begin{equation}
		\label{eq:averagePO2s}
		PO_{2,i}^{(t)} = \frac{1}{\left| CV_i \right|} \int_{CV_i} PO_2^{(t)}\;dV \text{ and }
		PO_{2,roi}^{(t)} = \frac{1}{\left| \Omega_{roi} \right|} \int_{\Omega_{roi}} PO_2^{(t)}\;dV.
		\end{equation}
		Having these values at hand, the following decisions are made: If $PO_{2,i}^{(t)}>36.5\;\text{mmHg}$, we stop the growth process in $CV_i$. By this an adaptive growth of the network is achieved. The whole growth process in this phase is stopped, if $PO_{2,roi}^{(t)}>36.5\;\text{mmHg}$. A threshold of $36.5\;\text{mmHg}$ is chosen, since in \cite[Table 2]{secomb2013angiogenesis} it is reported that the average $PO_2^{(t)}$ is around $32.5-35.5\;\text{mmHg}$. The threshold is chosen $2.5\;\text{mmHg}$ higher as the average $PO_2^{(t)} = 34.0\;\text{mmHg}$, due to the fact that in the third phase of the generation process some vessels are removed. In Algorithm \ref{alg:phase2} the main components of (P2) are summarised.
		\begin{algorithm}
			\caption{\label{alg:phase2} Algorithm for Phase 2}	
			\SetAlgoLined
			\textbf{INPUT:} 1D network $\Lambda_{P1}$ obtained from (P1), domain $\Omega$ \;
			Set $j = 0$, $\Lambda_j \leftarrow \Lambda_{P1}$, $PO_{2,roi,old}^{(t)} = 0$,
			StopPhase2 = \textbf{false}\;
			\While{!StopPhase2}{
				Determine the pressures $p_v$ and $p_t$ by solving \eqref{eq:3D1DTissue} in $\Omega$ and \eqref{eq:3D1DVascular} in $\Lambda_j$\;
				Compute the partial pressures $P_{O_2}^{(v)}$ and $P_{O_2}^{(t)}$  by solving \eqref{eq:3D1DTissue} in $\Omega$ and \eqref{eq:3D1DVascular} on	$\Lambda_j$\;
				Compute the local partial pressures $PO_{2,k}^{(t)},\;k \in \left\{1,\ldots,N_{cv}\right\}$ according to 
				\eqref{eq:averagePO2s}\;
				\For{all boundary nodes $x \in \Omega \cap \partial \Lambda_j $}{
					Determine $CV_k$ with $x \in CV_k$\;
					\eIf{$PO_{2,k}^{(t)}> 36.5 \;\unit{mmHg}$}{
						No new vessel is attached to $x$
					}{
						Construct new vessels as in Algorithm \ref{alg:phase1}, if the radii of the new vessels are below $3.0\;\mu m$ choose the new radii according to \eqref{eq:SmallRadii} \;
					}
				}
				Add the new vessels, try to link terminal vessels within the resulting network $\tilde{\Lambda}_j$ as shown in Figure \ref{fig:Connection} and update the network $\Lambda_{j+1} \leftarrow \tilde{\Lambda}_j$ \;
				Compute $PO_{2,roi,j}^{(t)}$ according to \eqref{eq:PO2_roi} \;
				\If{$\left|PO_{2,roi,j}^{(t)}-PO_{2,roi,old}^{(t)}\right|<1.0 \cdot 10^{-3}$ OR    $j>N_{it,\max}^{(2)}$ }{
					StopPhase2 = \textbf{true}\;
				}
				$PO_{2,roi,old}^{(t)} = PO_{2,roi,j}^{(t)}$, $\;j = j+1$ \;
			}
			\textbf{OUTPUT:} Network containing the main vessels and fine scaled vessels: $\Lambda_{P2} \leftarrow \Lambda_j$ \;
	\end{algorithm}	
	\item \textbf{Phase (P3):}
	\ \\
		In the third phase of the generation process, the network structure $\Lambda_{P2}$ obtained from (P2) is further optimised by removing dead ends i.e. terminal vessels contained in $\Omega_{roi}$. In addition to that, we try to connect terminal vessels with other vessels in the network. This is again executed in an iterative way by means of the following four steps: In the first step, we remove all the terminal vessels in $\Omega_{roi}$.
		\ \\ 
		After removing the terminal vessels, we try to connect the new terminal vessels with other vessels in the network as it is done in (P2). As in (P2), it is tested whether the new vessels are intersecting other vessels (see Algorithm \ref{alg:collision}). If less than $10$ terminal vessels can be found in $\Omega_{roi}$, we stop the growth process of the third phase. Finally we extract the network contained in $\Omega_{roi}$. A pseudo code for (P3) can be found in Algorithm \ref{alg:phase3}.
		\begin{algorithm}
		\caption{\label{alg:phase3} Algorithm for Phase 3}	
		\SetAlgoLined
		\textbf{INPUT:} 1D network $\Lambda_{P2}$ obtained from (P2), domain $\Omega$\;
		Set $j = 0$, $\Lambda_j \leftarrow \Lambda_{P2}$, $PO_{2,roi,old}^{(t)} = 0$,
		StopPhase3 = \textbf{false}\;
		Determine the number of terminal vessels of $\Lambda_j$ in $\Omega_{roi}$: $N_{term}$\;
		\While{!StopPhase3}{
			\For{all boundary nodes $x \in \Omega \cap \partial \Lambda_j $}{
				\If{$x \in \Omega_{roi}$}{
					Remove vessel $E_k^{(j)}$ containing $x$ from $\Lambda_j$\;
				}
			}
			Obtain an intermediate network $\tilde{\Lambda}_j$\;
			Try to link terminal vessels with other vessels in the network $\tilde{\Lambda}_j$ as shown in Figure \ref{fig:Connection}\;
			Update the network $\Lambda_{j+1} \leftarrow \tilde{\Lambda}_j$\;
			Determine the number of terminal vessels of $\Lambda_{j+1}$ in $\Omega_{roi}$: $N_{term}$\;
			\If{$j>N_{it,\max}^{(3)}$ OR $N_{term}<10$}{
				StopPhase3 = \textbf{true}\;
			}
			$j = j+1$\;
		}
		Extract network $\Lambda_{roi} = \Lambda_j \cap \Omega_{roi}$\;
		\textbf{OUTPUT:} Final network: $\Lambda_{P3} \leftarrow \Lambda_{roi}$\;
	\end{algorithm}
\end{itemize}

\subsection{Numerical methods}

Concluding the main section of this work, we briefly describe the numerical methods that have been used to determine a solution of the mathematical models developed in the previous subsections. To solve the 3D PDEs \eqref{eq:3D1DTissue} and \eqref{eq:TissuePO23D1D}, we use a standard cell-centered finite volume method \cite{helmig1997multiphase}. The fluxes across the boundaries of a finite volume cell are approximated by the two-point flux method. Using a simple two-point approximation of the fluxes yields a consistent discretisation, since a simple uniform hexahedral mesh decomposing $\Omega$ is used. Moreover, we have assumed that the tissue is an isotropic porous medium and that the diffusion coefficient for oxygen in tissue is constant. To incorporate the Dirac source term in the 3D equations \eqref{eq:3D1DTissue} and \eqref{eq:TissuePO23D1D} using finite volumes, we first determine which finite volume cells are intersected by the cylindrical vessels. In a next step, the areas of vessel surfaces contained in the affected finite volume cells are estimated. Thereby, it is crucial that the area of a cylindrical vessel is consistently distributed among the respective finite volume cells. This means that the sum of the single surface areas assigned to the 3D finite volume cells is equal to the surface area of the respective cylinder. By this feature, we obtain a mass conservative discretisation of the flow and transport models. An advantage of this numerical solution method is that the 1D graph structure governing the location of the vascular network does not have to be aligned with the edges of the 3D finite volume mesh as e.g. in \cite{kojic2017composite} the authors require that the edges of the 3D mesh are adapted to the edges of the vascular graph.

For the numerical solutions of the 1D PDEs \eqref{eq:3D1DVascular} and \eqref{eq:VascularPO23D1D}, we employ the vascular graph model (VGM) \cite{erbertseder2012coupled, peyrounette2018multiscale,reichold2009vascular,vidotto2019hybrid}, which corresponds in principle to a vertex centered Finite Volume method with a two-point flux approximation. The discretised versions of the 3D and 1D PDEs are linear in the fluid pressures and partial pressures of oxygen. However, there is one exception caused by the source term of \eqref{eq:TissuePO23D1D}. Since this source term involves the Michaelis-Menten law \eqref{eq:MichaelisMenten}, a nonlinearity is introduced and a nonlinear solver is required. For our simulations, we choose a simple fixed point iteration with a damping factor to enforce the convergence of the scheme. As an initial guess for the fixed point iteration, we use the distribution of partial pressure of oxygen from the previous growth step. 

\section{Simulation results}
\label{sec:SimulationResults}

After describing a numerical model for generating surrogate networks, we test the performance of this numerical model. This is done by comparing some important features of the segmented network shown in Figure \ref{fig:networks} (left) with the corresponding features of the created network. In particular, the total length 
$L = \sum_{k \in I} \left\| E_k \right\|_2$, the total vessel surface area $A = \sum_{k \in I} \left\| E_k \right\|_2 2 R_k \pi$, the total volume $V = \sum_{k \in I} \left\| E_k \right\|_2 R_k^2 \pi$ and the number of segments $N_{seg} = \left|I \right|$ of a network $\Lambda = \bigcup_{k \in I} E_k$ are considered. $I$ is an index set numbering the segments of $\Lambda$. The reason $L$ is reported, is due to the fact that the surrogate network should exhibit a similar flow path as the segmented network. If the surface area $A$ is in both networks in the same range, it can be expected that under similar boundary conditions the fluid exchange between the tissue and the vascular system is in both cases in the same range. The volume $V$ is a quantity of interest, since networks having a similar volume contain a similar amount of blood volume. In addition to that $PO_{2,roi}^{(t)}$ and an averaged fluid pressure in tissue $p_{t,roi}$:
$$
p_{t,roi} = \frac{1}{\left| \Omega_{roi} \right|} \int_{\Omega_{roi}} p_t\;dV
$$
are presented. Furthermore, the flux $F_{tv}\;\left[ \unitfrac{\mu g}{s} \right]$ from the vascular system into the tissue and vice versa is calculated. Therefore, we multiply the volumetric flux by the density of water $\left(\rho_w = 1000.0 \;\unitfrac{kg}{m^3}\right)$. Finally, we take the total number of growth steps $N_{it}$ for all the three growth phases into account. All these data are reported varying two model parameters that affect the shape of the generated networks in a significant way. These parameters are the consumption rate of oxygen $m_0^{(ox)}$ and the Murray parameter $\gamma$. The rest of this section is divided into two parts. In the first part, we list all the model parameters that are used for the simulations, while in the second part, the simulation results are presented and discussed.

\subsection{Simulation parameters}
\label{sec:Simulation_parameters}

\begin{figure}[h!]
	\begin{center}		
		\includegraphics[width=0.485\textwidth]{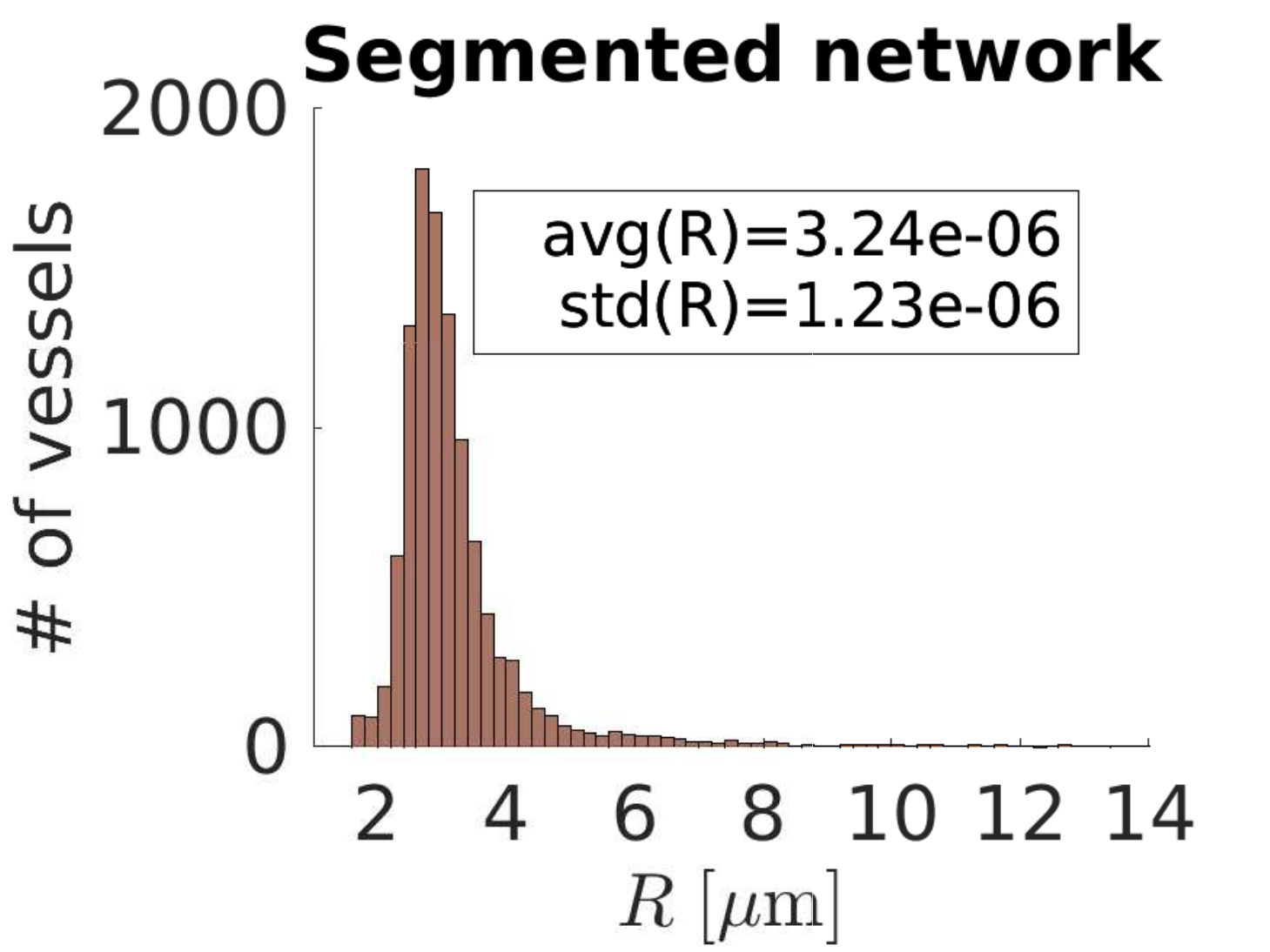}
		\includegraphics[width=0.495\textwidth]{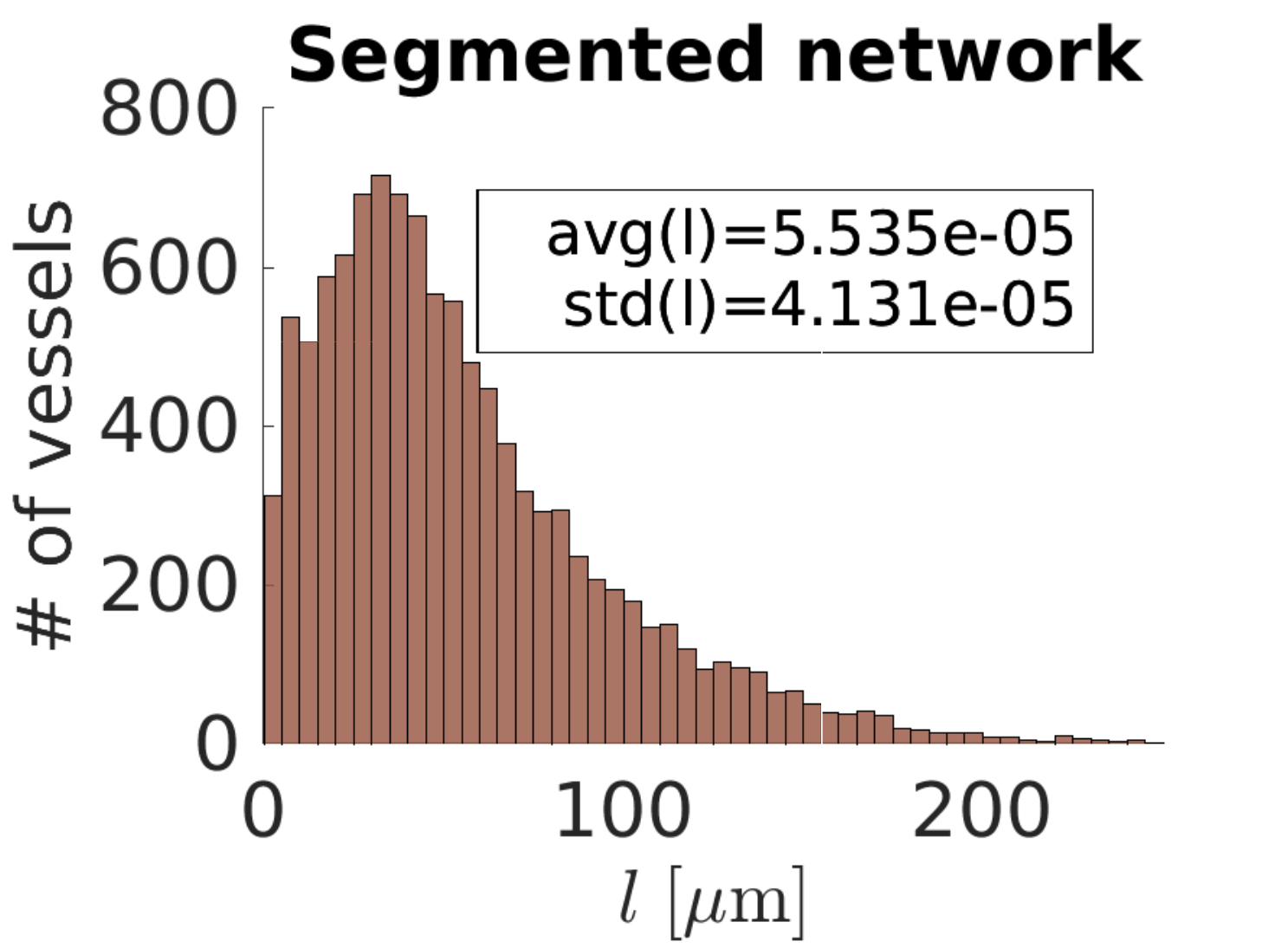}
	\end{center}
	\caption{\label{fig:segmented} This figure shows two histogramms with respect to the segmented network shown in Figure \ref{fig:networks}. On the left hand side, the distribution of the radii is shown, while on the right hand side the distribution of the lengths is presented. The terms avg and std stand for the mean value and standard deviation, respectively.}
\end{figure}

Figure \ref{fig:segmented} shows the distribution of radii and lengths of the segments that are part of the given network $\Lambda_s$ shown in Figure \ref{fig:networks} (left). By means of these distributions, we compute the corresponding length $L_s$, surface area $A_s$, volume $V_s$ and the total edge number $N_s$. The different values can be found in Table \ref{tab:values_segmented}.
\begin{table}[h!]
	\centering
	\caption{\label{tab:values_segmented} Characteristic features of the segmented network (see Figure \ref{fig:networks}, left).}
	\begin{center}
		\begin{tabular}{|c|c|c|c|}
			\hline
			\hline
			Name & 	Sign & Value & Unit \\
			\hline
			\hline
			total length & $L_s$ & $0.595$ & $\unit{m}$ \\
			\hline
			total surface area & $A_s$ & $1.17 \cdot 10^{-5}$ & $\unit{m}^2$ \\
			\hline
			total volume & $V_s$ & $2.21\cdot 10^{-11}$ & $\unit{m}^3$ \\
			\hline
			total number of segments & $N_{s,seg}$ & $10746$ & $\unit{-}$ \\
			\hline
			\hline
		\end{tabular}
	\end{center}				
\end{table}
The next table contains all the remaining parameters occurring in the 3D-1D coupled PDE based models \eqref{eq:3D1DVascular}, \eqref{eq:3D1DTissue} and \eqref{eq:VascularPO23D1D}, \eqref{eq:TissuePO23D1D}. It can be seen at a first glance that most of the parameters are fixed apart from $m_0^{(ox)}$ and $\gamma$. We have chosen 
$\gamma \in \left\{3.0,3.5\right\}$, since these are usual values for the Murray parameter $\gamma$ one can find in literature \cite{fung1997biomechanics,murray1926physiological,schneider2012tissue}. In \cite[Table 1]{secomb2013angiogenesis} the authors provide for the oxygen consumption rate $c_0^{(ox)}$ a range of $0.5\;\unit{cm^3\;O_2\;\left(100\;cm^3\right)^{-1}\;min^{-1}}$ to $2.5\;\unit{cm^3\;O_2\;\left(100\;cm^3\right)^{-1}\;min^{-1}}$. By means of Henry's law $m_0^{(ox)}=c_0^{(ox)}/\alpha_H$ and the Henry constant $\alpha_H = 2.894 \cdot 10^{-5}\;\unit{cm^3\;O_2\;cm^{-3}\;mmHg^{-1}}$ \cite[Table 4]{koppl2014influence}, we obtain:
$m_0^{(ox)} \in \left[ 2.88\;\unitfrac{mmHg}{s}, 14.40\;\unitfrac{mmHg}{s} \right]$. Due to the fact that a portion of rat brain is considered, it can be assumed that low oxygen consumption rates should be used for the simulations. Therefore, we have chosen for $m_0^{(ox)}$ the values $3.00\;\unitfrac{mmHg}{s}$ and $4.00\;\unitfrac{mmHg}{s}$.
{\small
\begin{table}[h!]
	\centering
	\caption{\label{tab:model_parameters} Model parameters of the 3D-1D coupled models for blood flow and oxygen transport.}
	\begin{center}
		\begin{tabular}{|l|c|c|c|c|}
			\hline
			\hline
			Name & 	Sign & Value & Unit & Source \\
			\hline
			\hline
			Permeability of tissue & $K_t$ & $1.0 \cdot 10^{-18}$ & $\unit{m}^2$ & \cite[Sec. 4.5]{cattaneo2014fem}	\\
			\hline
			Permeability (plasma) & $L_p$ & $1.0 \cdot 10^{-12}$ & $\unitfrac{m}{Pa \cdot s}$ & \cite[Sec. 4.5]{cattaneo2014fem}	\\
			\hline
			Viscosity blood plasma & $\mu_p$ & $1.0 \cdot 10^{-3}$ & $\unit{Pa \cdot s}$ & \cite{pries1996biophysical} \\
			\hline
			Reflection parameter & $\sigma$ & $0.1$ &  $\unit{-}$ & $\unit{-}$ \\
			\hline
			Onc. pressure blood plasma & $\pi_v$ & $3733.0$ & $\unit{Pa}$ & \cite[Tab. 5.1]{cattaneo2014fem} \\
			\hline 
			Onc. pressure interstital fluid & $\pi_t$ & $666.0$ & $\unit{Pa}$ & \cite[Tab. 5.1]{cattaneo2014fem} \\
			\hline
			Viscosity interstital fluid & $\mu_t$ & $1.30 \cdot 10^{-3}$ & $\unit{Pa \cdot s}$ & \cite[Tab. 5.1]{cattaneo2014fem} \\
			\hline
			Diff. of oxygen (vascular) & $D_v$  & $5.00 \cdot 10^{-5}$ & $\unitfrac{m^2}{s}$ & \cite[Tab. 5.2]{cattaneo2014fem} \\  
			\hline
			Diff. of oxygen(tissue) & $D_t$  & $1.35 \cdot 10^{-7}$ & $\unitfrac{m^2}{s}$ & \cite[Tab. 5.2]{cattaneo2014fem} \\  
			\hline
			Metabolic rate & $m_0^{(ox)}$ & $3.00$, $4.00$ & $\unitfrac{mmHg}{s}$ & \cite[Tab. 1]{secomb2013angiogenesis} \\
			\hline
			$P_{O_2}$ (half consump.) & $P_{O_2,0}^{(t)}$ & $1.00$ & $\unit{mmHg}$ & \cite[Tab. 1]{secomb2013angiogenesis}\\
			\hline
			Permeability (oxygen) & $L_{P_{O_2}}$ & $3.50 \cdot 10^{-5}$ & 
			$\unitfrac{m}{s}$ & \cite[Table 5.2]{cattaneo2014fem} \\  
			\hline
			Murray parameter & $\gamma$ & $3.00,\;3.50$ & $\unit{-}$ & \cite[Tab. 1]{schneider2012tissue} \\
			\hline
			Regularisation parameter & $\lambda_g$ & $1.0$ & $\unit{-}$ & \cite[Tab. 1]{schneider2012tissue} \\
			\hline
			Lower bound $\Omega_{roi},\;x_1$ & $L_{1,l}^r$ & $0.038$ & $\unit{mm}$ & $\unit{-}$ \\
			\hline
			Upper bound $\Omega_{roi},\;x_1$ & $L_{1,u}^r$ & $1.13$ & $\unit{mm}$ & $\unit{-}$  \\
			\hline
			Lower bound $\Omega_{roi},\;x_2$ & $L_{2,l}^r$ & $8.8 \cdot 10^{-4}$ & $\unit{mm}$ & $\unit{-}$ \\
			\hline
			Upper bound $\Omega_{roi},\;x_2$ & $L_{2,u}^r$ & $1.05$ & $\unit{mm}$ & $\unit{-}$ \\
			\hline
			Lower bound $\Omega_{roi},\;x_3$ & $L_{3,l}^r$ & $8.8 \cdot 10^{-4}$ & $\unit{mm}$ & $\unit{-}$ \\
			\hline
			Upper bound $\Omega_{roi},\;x_3$ & $L_{3,u}^r$ & $1.50$ & $\unit{mm}$ & $\unit{-}$ \\
			\hline
			\hline		
		\end{tabular}
	\end{center}				
\end{table}}	
The information on the network $\Lambda_{lv}$ containing the larger and well segmented vessels can be found in the file \emph{extracted\_network.dgf}, which is uploaded together with the supplementary material. In this file the nodes of the graph-like structure are listed. Next to the coordinates of the boundary nodes a blood pressure in $\left[ \unit{Pa} \right]$ is listed. These values are used as boundary conditions for the 1D PDEs \eqref{eq:3D1DVascular}. In addition to that the segments of $\Lambda_{lv}$ are defined i.e. there is a list consisting of three columns, where the first column contains the index of the first node and the second column the index of the second node. Finally, the third column presents the radius of the corresponding segment in $\left[ \unit{m} \right]$. It remains to specify the boundary conditions for the partial pressure of oxygen $PO_2^{(v)}$. According to \cite{secomb2013angiogenesis}, the $PO_2^{(v)}$ in arterial vessels is around $75.0\;\unit{mmHg}$ and in venous vessels is around $38.0\;\unit{mmHg}$. In order to identify the arterial and venous blood vessels, we compute first the flow field in $\Lambda_{lv}$ and the average velocity for each edge. If an edge exhibits a velocity below the average velocity, we regard it as a vein. Otherwise, it is considered as an artery. 

\subsection{Results and discussion}
\label{sec:results}

\begin{figure}[h!]
	\includegraphics[width=0.5\textwidth]{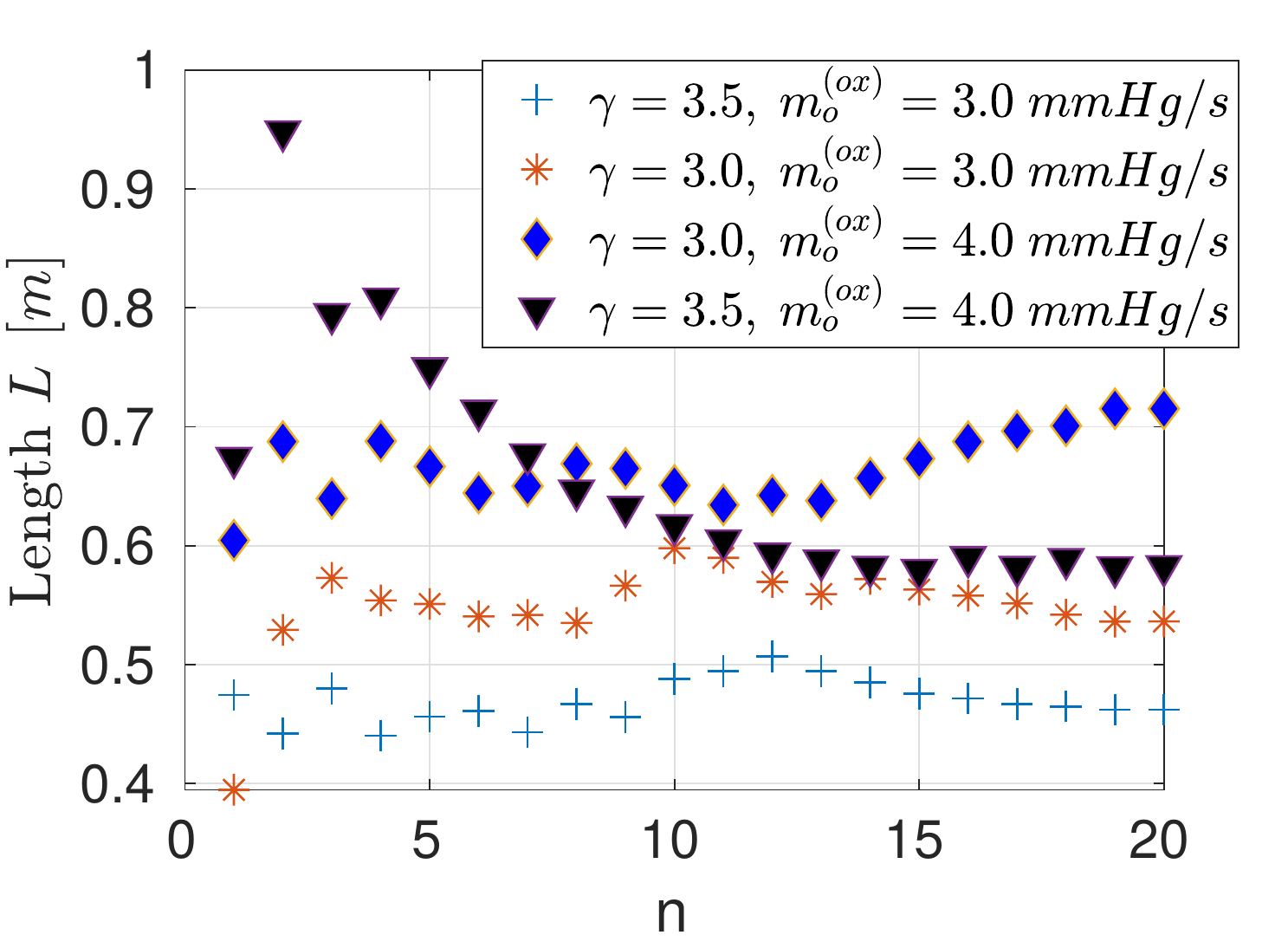}
	\includegraphics[width=0.5\textwidth]{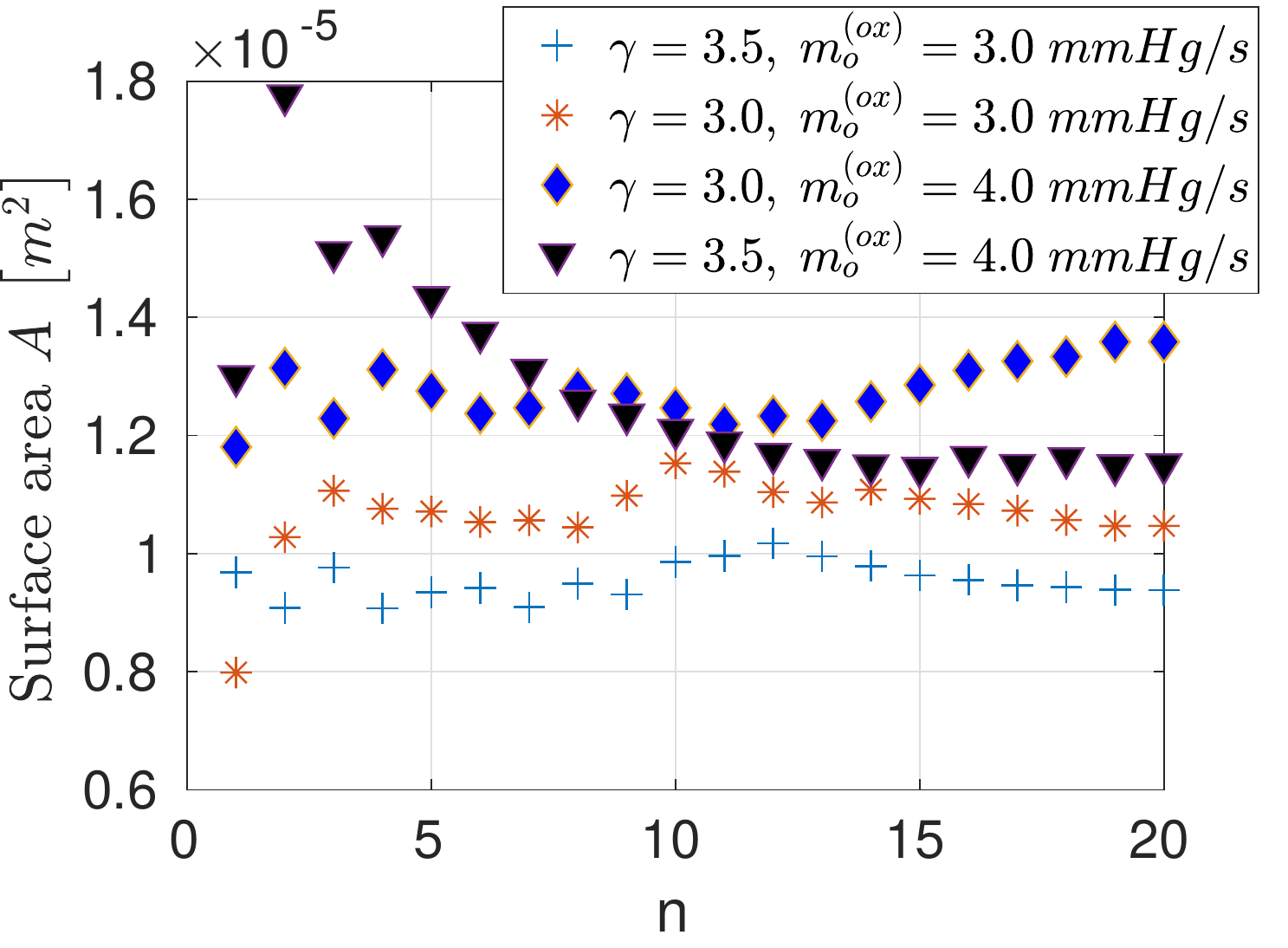}
	\includegraphics[width=0.5\textwidth]{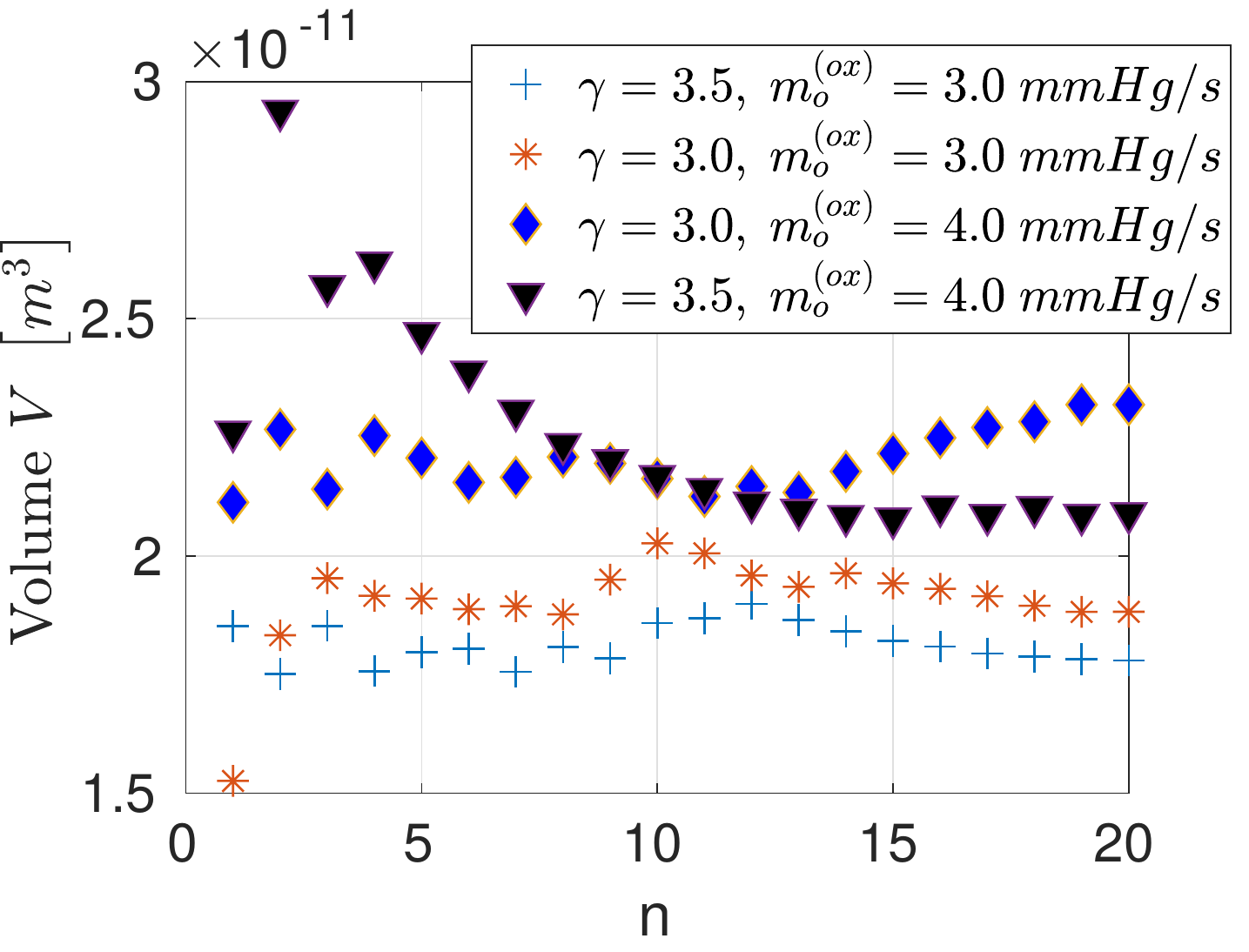}
	\includegraphics[width=0.5\textwidth]{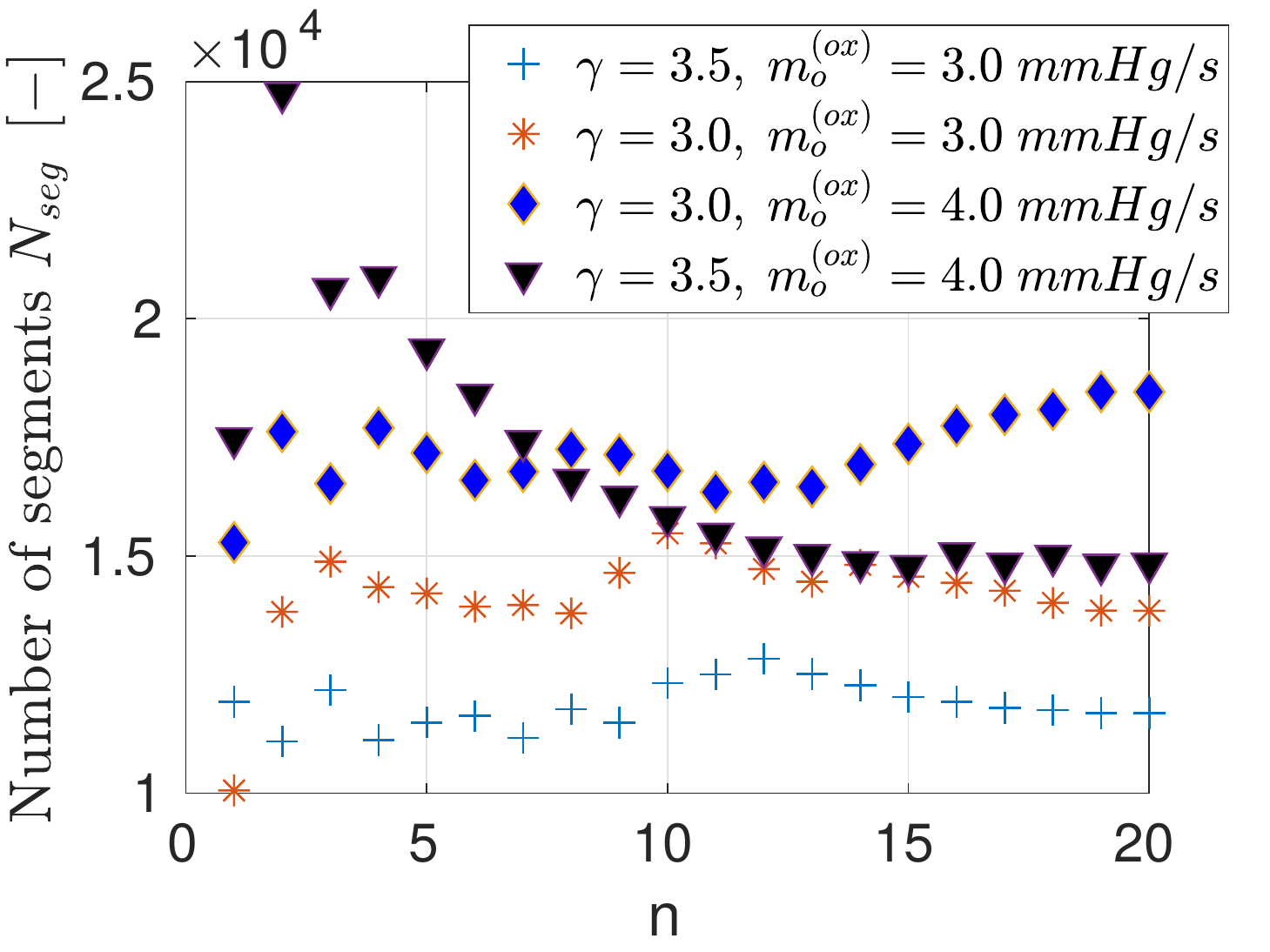}
	\caption{\label{fig:asymptotic_1} Asymptotic behaviour of the total length $L$, total surface area $A$, total volume $V$ and the total number of segments $N_{seg}$.}
\end{figure}

\begin{figure}[h!]
	\includegraphics[width=0.5\textwidth]{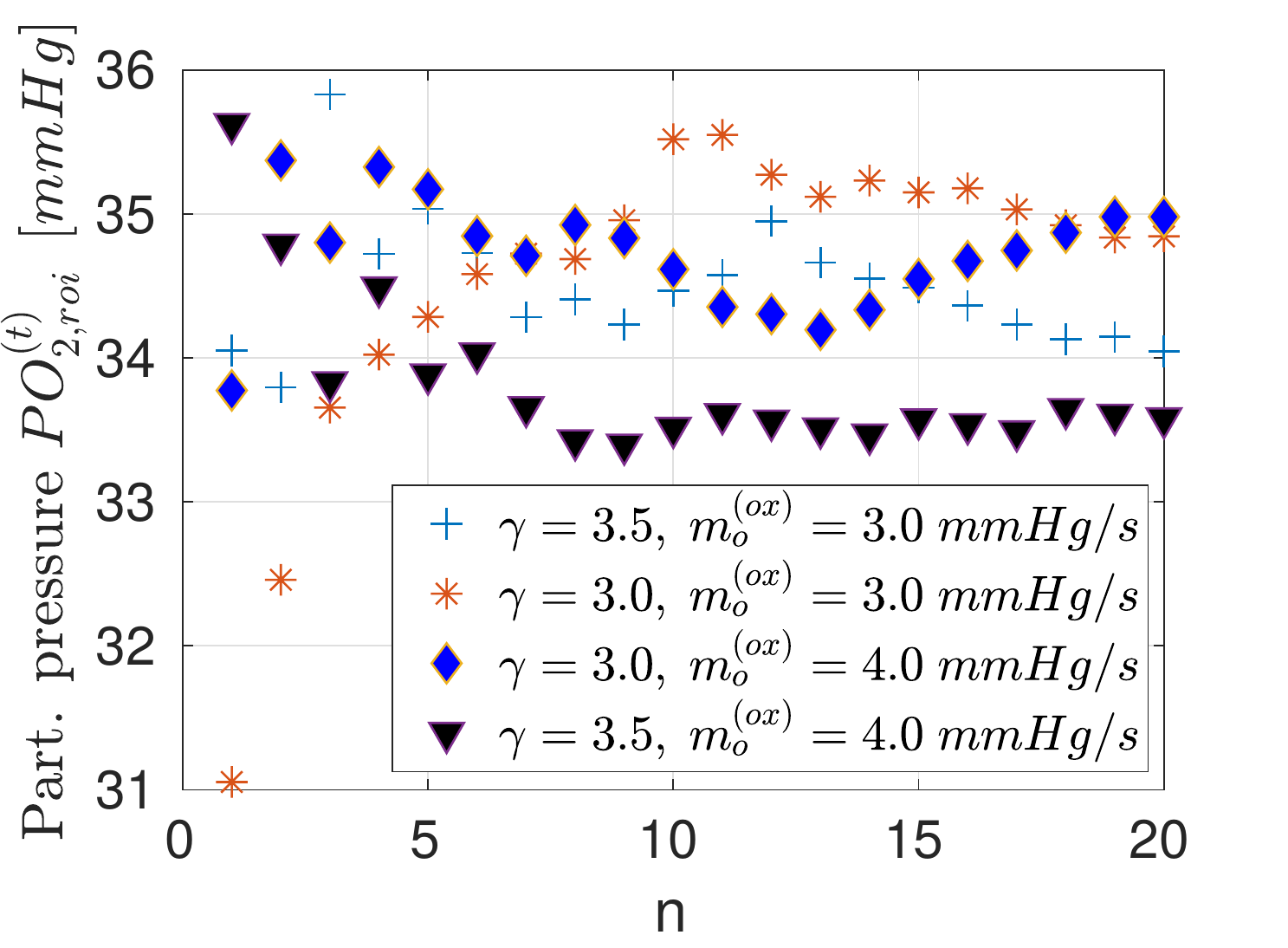}
	\includegraphics[width=0.5\textwidth]{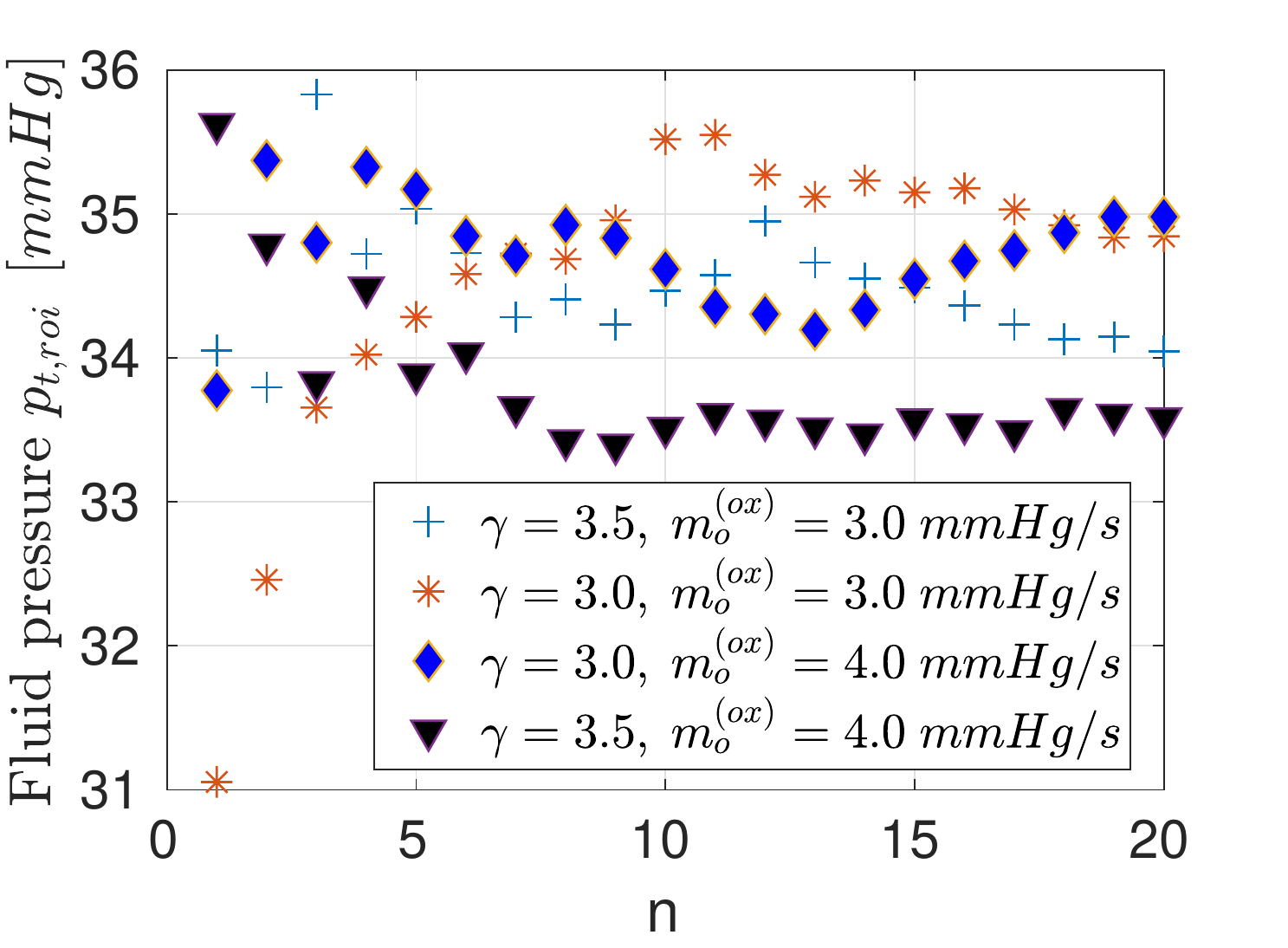}
	\includegraphics[width=0.5\textwidth]{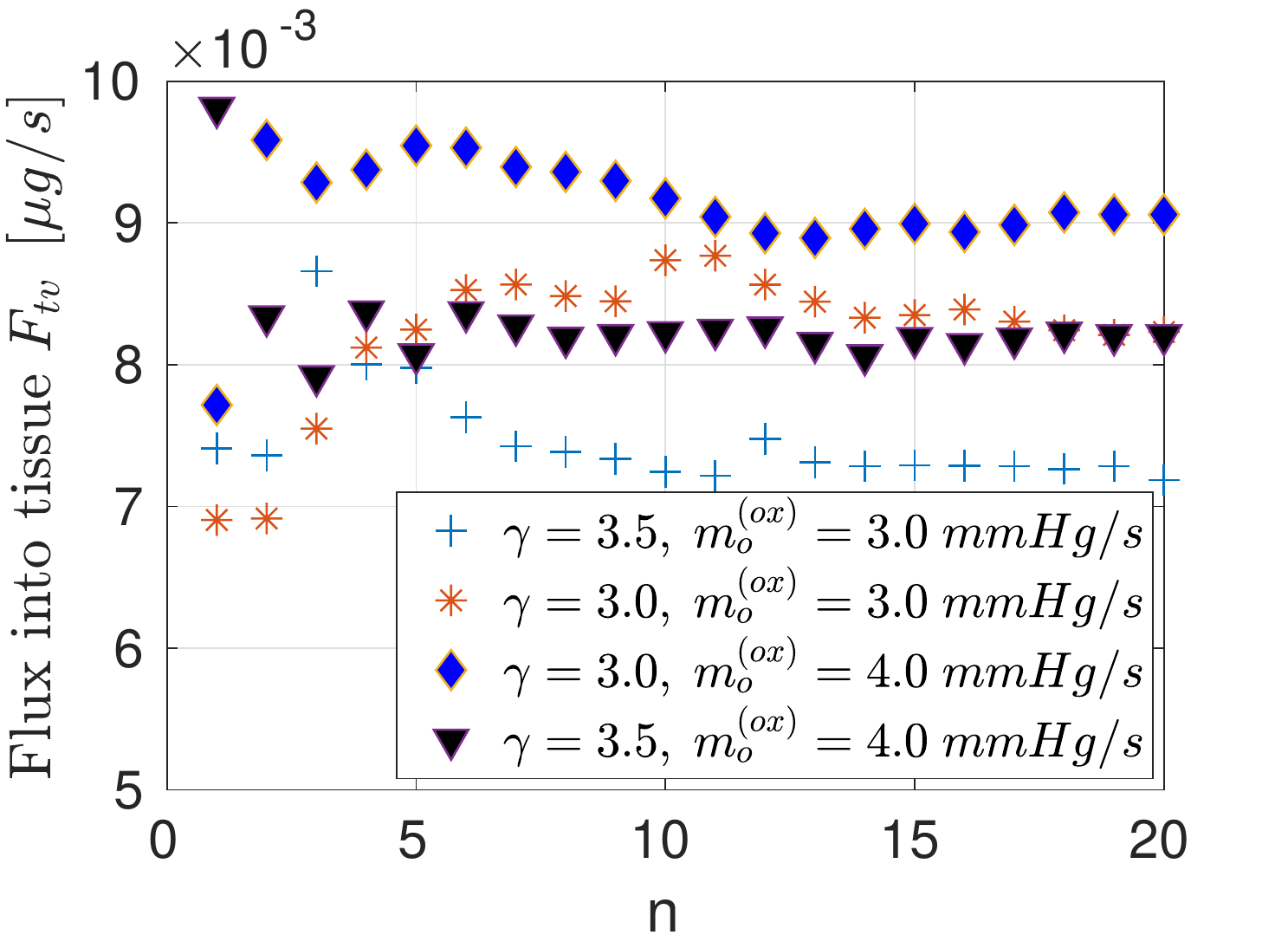}
	\includegraphics[width=0.5\textwidth]{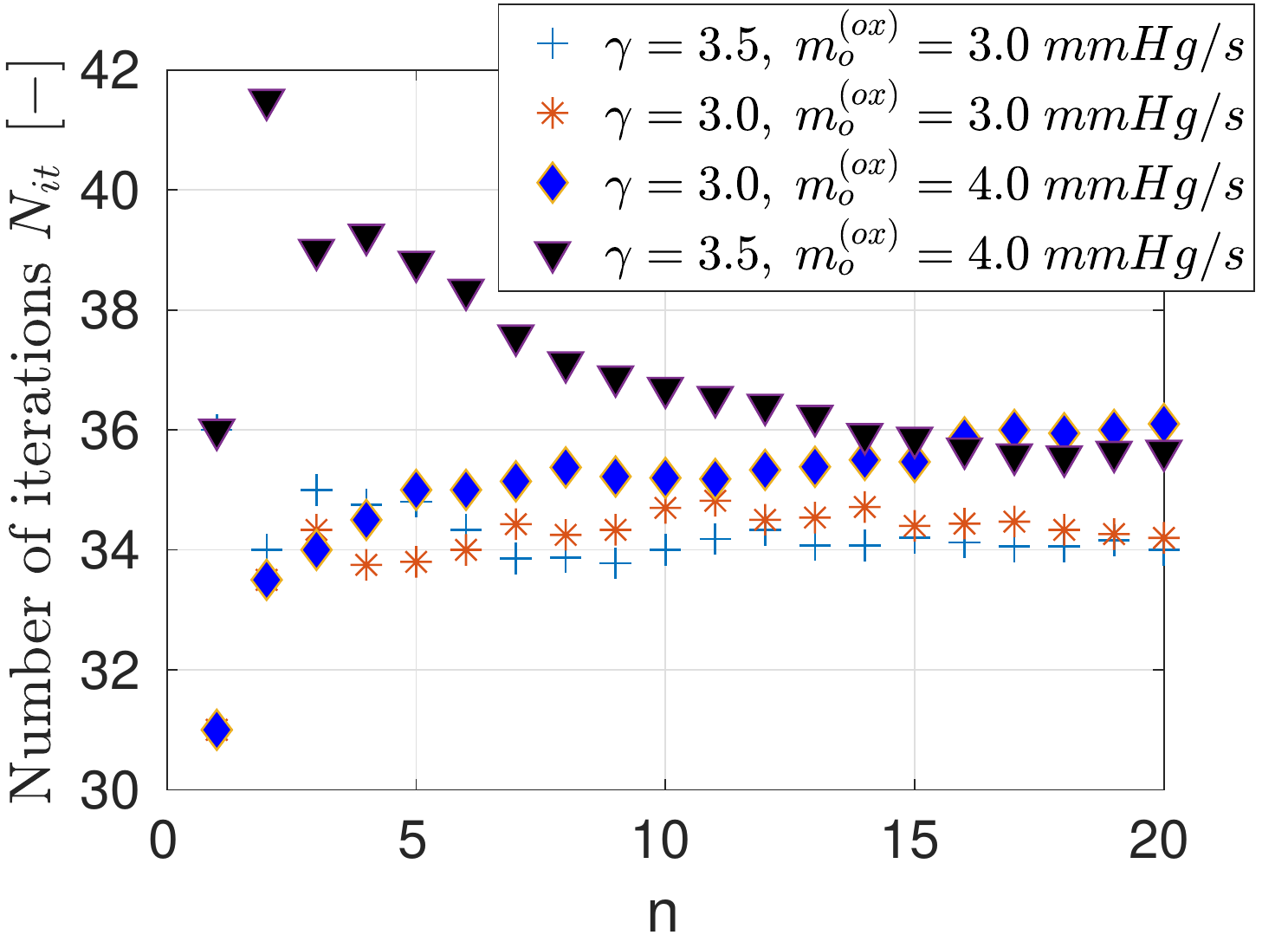}
	\caption{\label{fig:asymptotic_2} Asymptotic behaviour of the pressures $PO_{2,roi}^{(t)}$, $p_{t,roi}$, the flux into tissue $F_{tv}$ and the total number of growth steps $N_{it}$.}
\end{figure}

At first, we investigate, how many simulations are required to obtain, for each parameter choice, a representative data set. This is necessary, since the growth process depends on several stochastic processes e.g. the formation of bifurcations. For this purpose, we perform $N_m\left( \gamma,m_0^{(ox)}\right)$ simulations and report all the quantities of interest that are listed at the beginning of this section. Then, for each quantity 
$q \in \left\{ L,\;A,\;V,\;N_{seg},\;PO_{2,roi}^{(t)},\;p_{t,roi},\;F_{tv},\;N_{it}\right\}$, the following mean values $q_{m_i}$ are computed:
$$
q_{m_i} = \frac{1}{i} \sum_{n=1}^i q_n\left( \gamma,m_0^{(ox)}\right),\;1 \leq i \leq N_m\left( \gamma,m_0^{(ox)}\right),
$$
where $q_n\left( \gamma,m_0^{(ox)}\right)$ denotes the quantity $q$ for a certain data set and the $n$-th simulation. As it can be observed numerically (see Figures \ref{fig:asymptotic_1} and \ref{fig:asymptotic_2}), a small number of samples is sufficient to find a quite good approximation of the quantities of interest. More precisely with only $N_m\left( \gamma,m_0^{(ox)}\right)=20$, we get for almost all cases satisfying results. Accordingly, we can assume that in this case that we have a representative data set. In Table \ref{tab:means_deviations}, the mean values and standard deviations for the different parameter pairs are listed. Comparing these data with the ones in Table \ref{tab:values_segmented}, we see that the quantities which determine the morphology of the network are closest to the segmented ones for $\gamma =3$ and $m_0^{(ox)} = 3.0\;\unitfrac{mmHg}{s}$. All the other data sets deviate to a larger extend from the segmented data. 

\begin{figure}[h!]
	\includegraphics[width=0.5\textwidth]{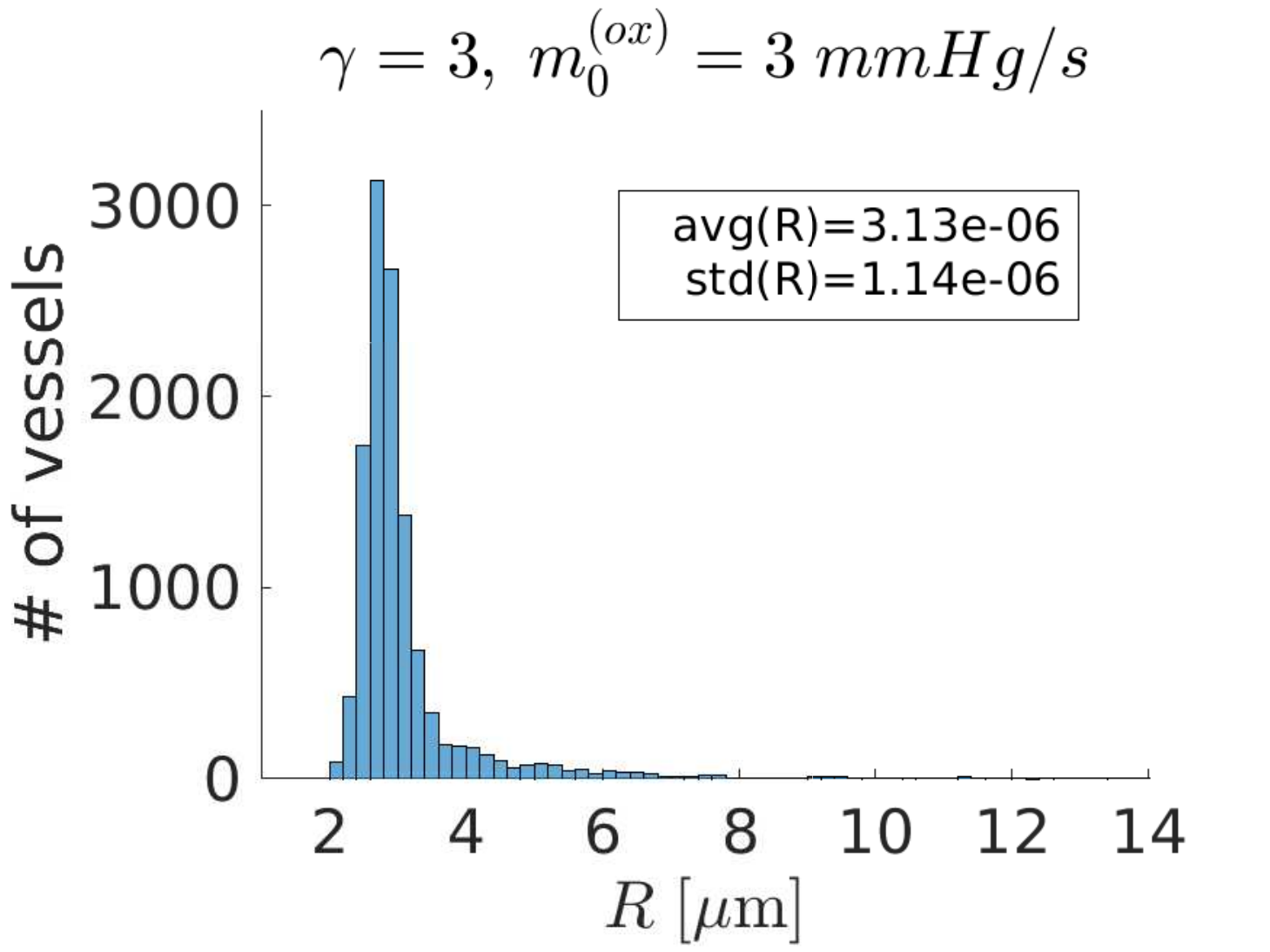}
	\includegraphics[width=0.5\textwidth]{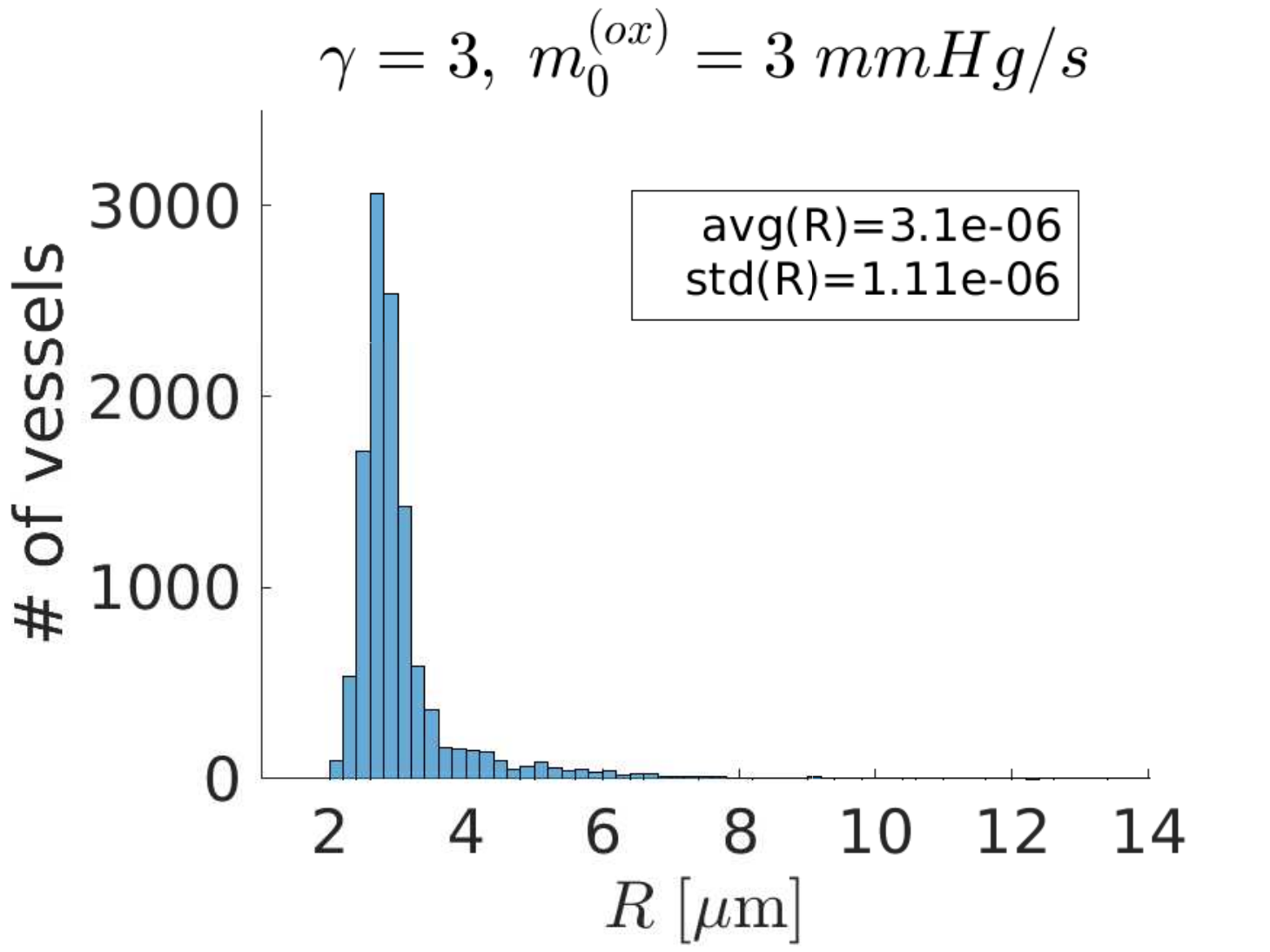}
	\includegraphics[width=0.5\textwidth]{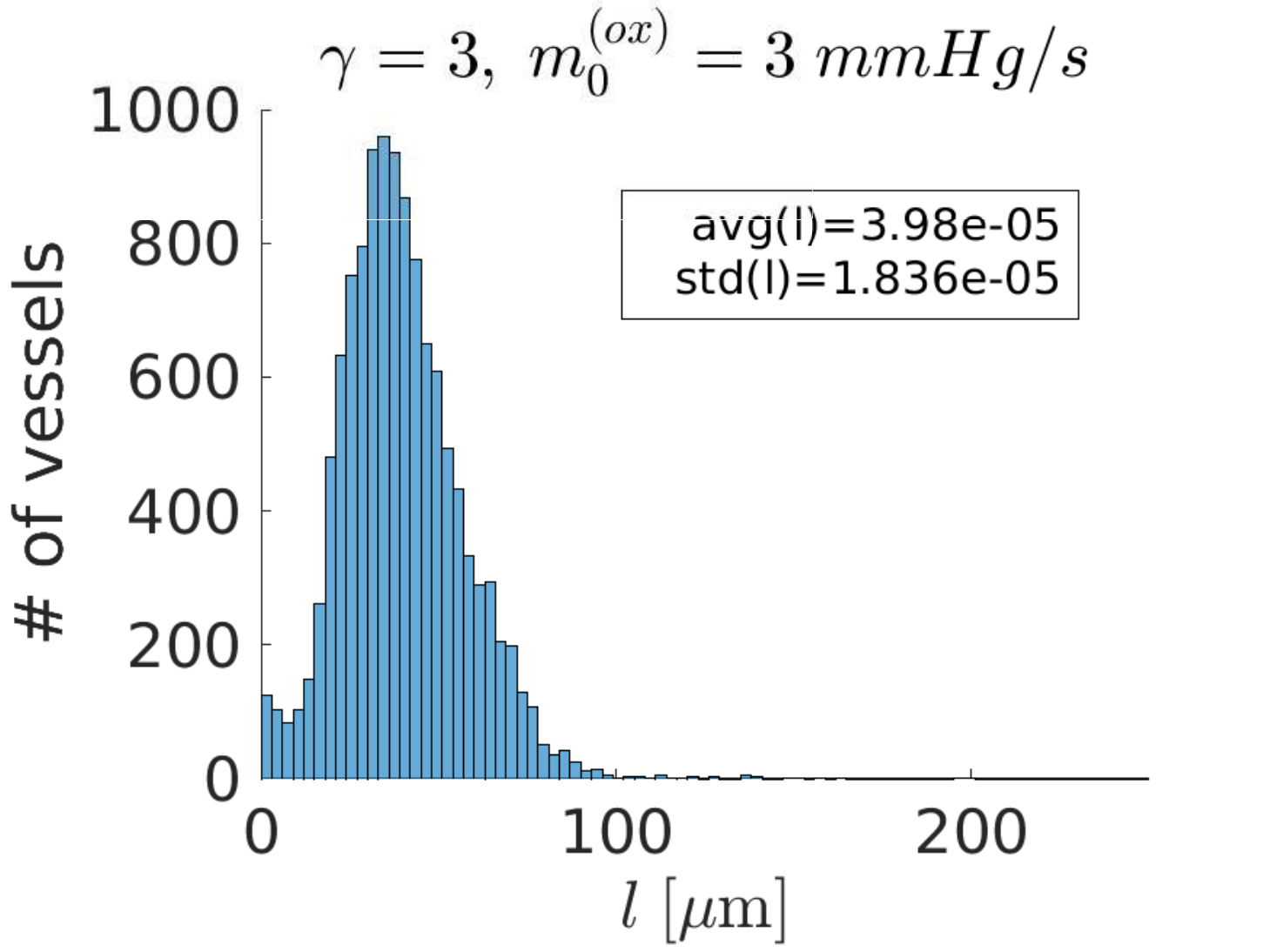}
	\includegraphics[width=0.5\textwidth]{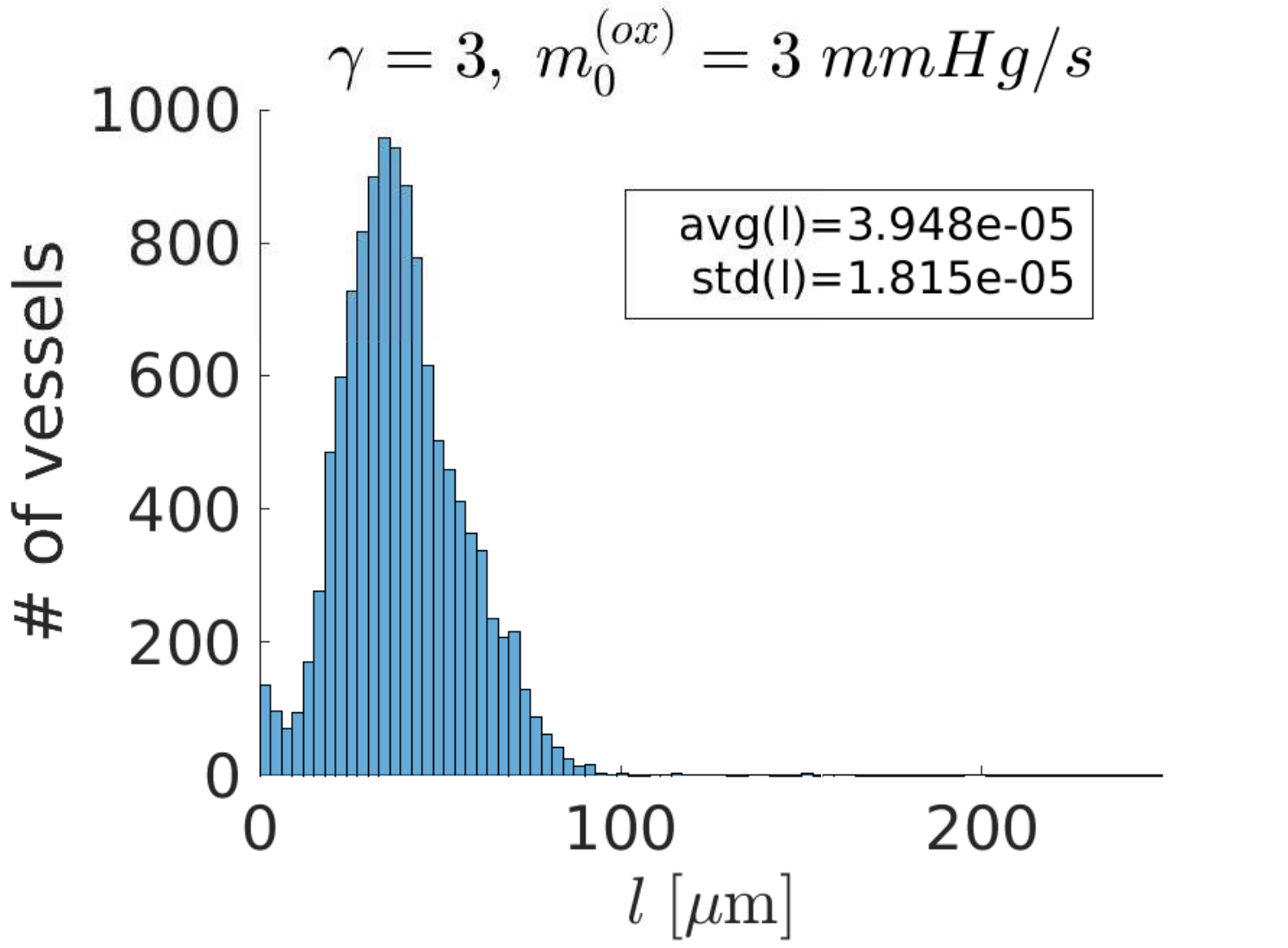}
	\caption{\label{fig:distributions_artificial} Two distributions of radii and segment lengths for $\gamma =3$ and
		$m_0^{(ox)}=3.0\;\unitfrac{mmHg}{s}$.}
\end{figure}

Next we compare for this parameter set two distributions of radii and lengths with their segmented counterparts (see Figure \ref{fig:segmented} and Figure \ref{fig:distributions_artificial}). It can be seen that the distributions of the radii have a similar shape as the one of the segmented data and that the mean value as well as the standard deviation are relatively close to each other. The main difference in this context is that more vessels are distributed in a small vicinity of the mean value. Contrary to that the distributions of the edge lengths exhibit different behaviours. While in the surrogate case, almost all segments have lengths between $1.0\; \mu\unit{m}$ and $100.0\; \mu\unit{m}$, the length distribution in the segmented case is decaying much slower, which means that a significant amount of the edges has a length of more than $100.0\; \mu\unit{m}$. This difference might be explained by the fact that an edge length in the segmented network does not necessarily correspond to a biological blood vessel, whereas in the other case we try to adapt the edge length to a typical blood vessel length. However, the total length of the networks is quite similar due to the fact that the surrogate network contains more vessels than the segmented one (see Table \ref{tab:values_segmented} and \ref{tab:means_deviations}). 

\begin{figure}[h!]
	\begin{center}
		\includegraphics[width=1.0\textwidth]{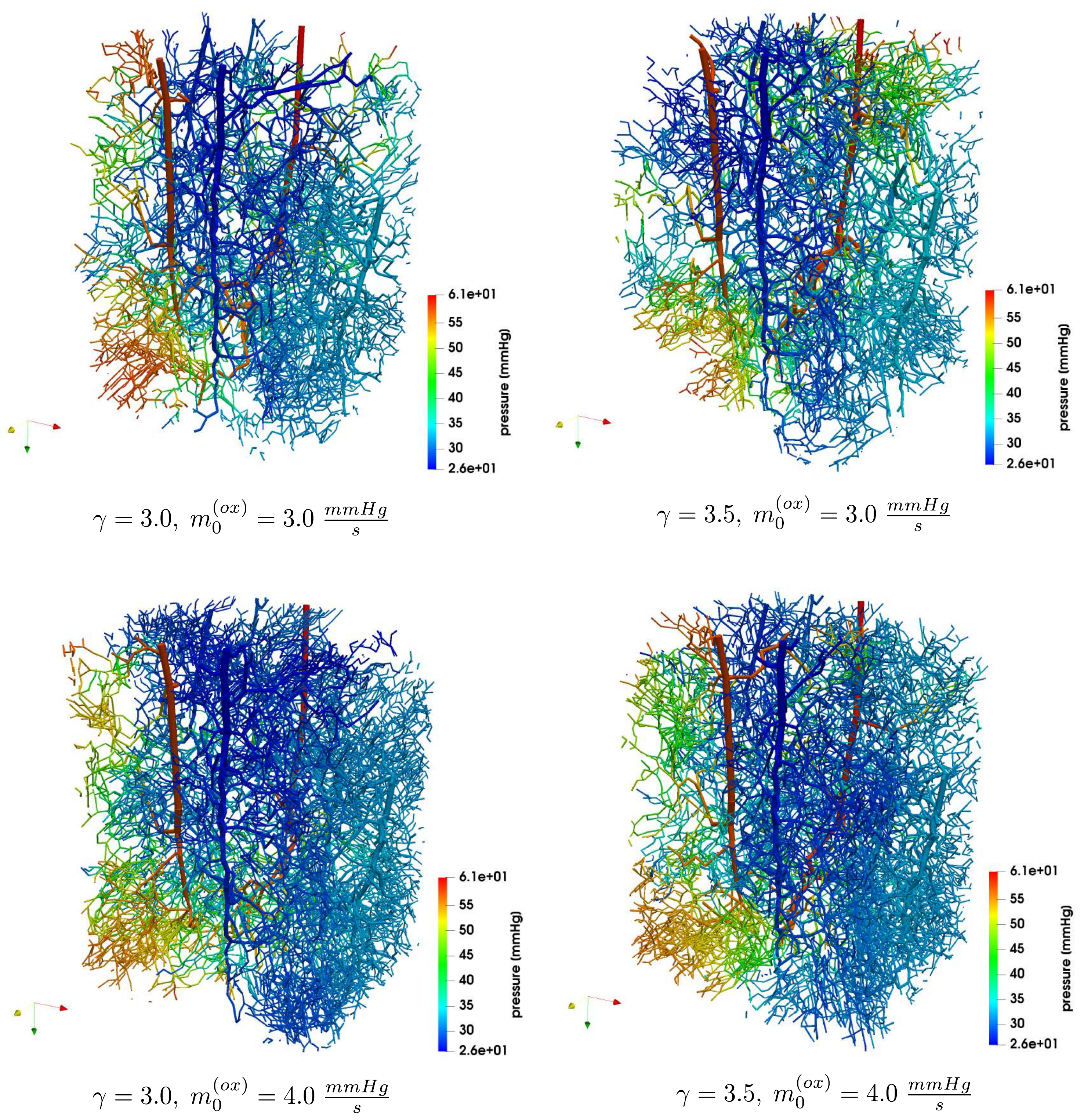}
	\end{center}
	\caption{\label{fig:extracted_networks} This figure shows examples of four networks contained in $\Omega_{roi}$ and obtained after the third phase. All the depicted networks are simulated by means of different parameters and represent the typical results for the respective data set. The colouring of the networks is given by the blood flow pressure ranging from $26\;\unit{mmHg}$ to $61\;\unit{mmHg}$.}
\end{figure}

\begin{figure}[h!]
	\begin{center}
		\includegraphics[width=1.0\textwidth]{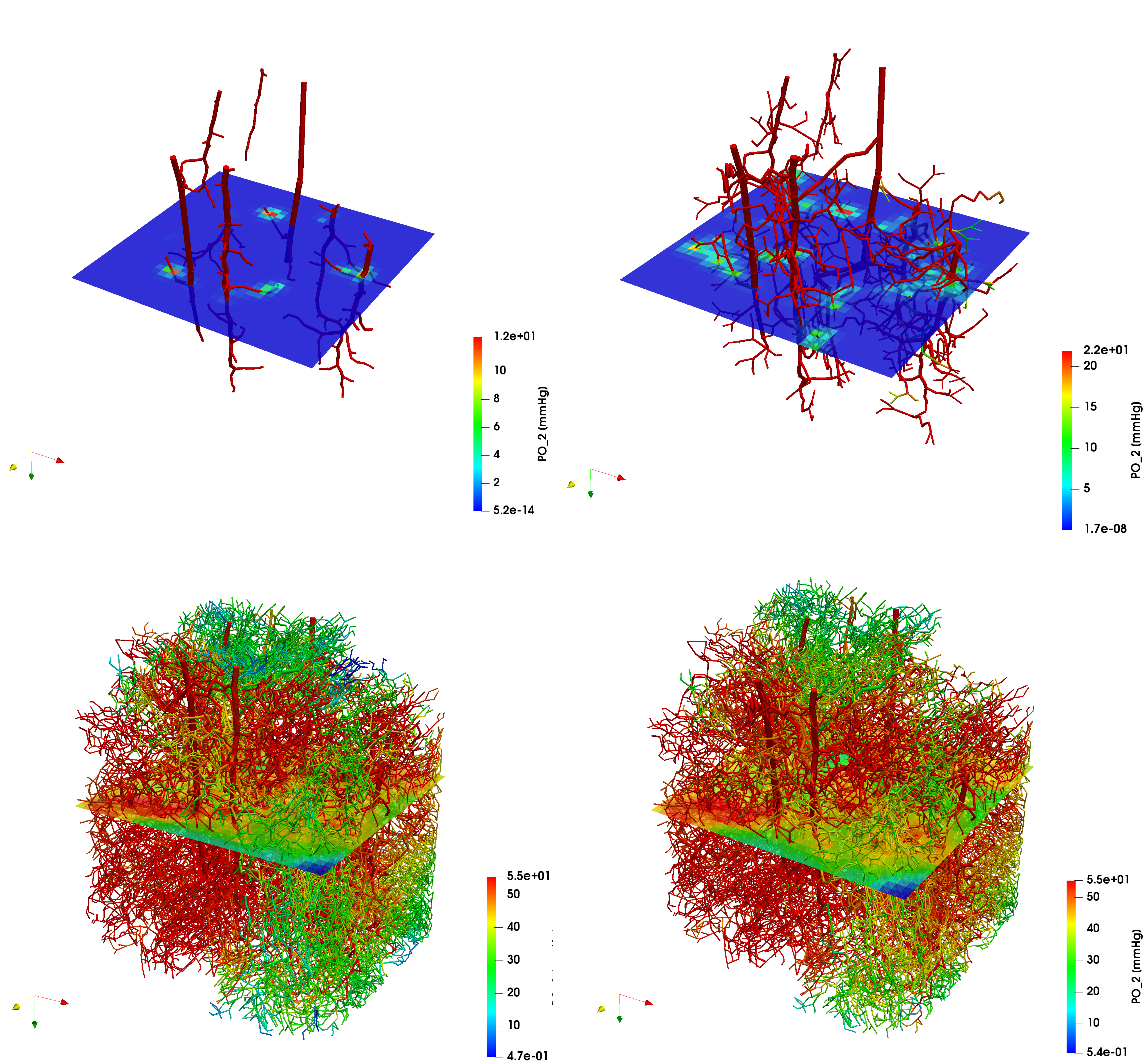}
	\end{center}
	\caption{\label{fig:growth_phases} In the top left picture of this figure the starting network for vessel growth is shown, while the top right picture shows the network after the first growth phase. At the bottom the networks after the second phase (left) and third phase (right) is presented. Each picture shows a tissue slice. The colour bar is adapted to the $PO_2$ within the tissue slice.}
\end{figure}

Finally, we illustrate in Figures \ref{fig:extracted_networks} and \ref{fig:growth_phases}, how the different parameters affect the shape of the networks and how the networks contained in $\Omega$ typically look like after the different phases. The networks become denser, if the rate of oxygen consumption is higher. This is due to the fact that for a higher consumption rate the propagation front extends across a much shorter distance than for a lower consumption rate. In order to compensate the reduced distance and maintain an average $PO_2^{(t)}$ of about $33\;\unit{mmHg}-35\;\unit{mmHg}$, the network has to become denser. As a consequence, the total surface area $A$ of the vessels and the total volume $V$ increases, which implies that the flux into tissue as well as the blood volume in the vascular system is increased. However, constructing the denser networks requires more growth steps $N_{it}$ compared to the low consumption case (see Table \ref{tab:means_deviations}). The blood pressure in the network ranges between $61.0\;\unit{mmHg}$ (arterial pressure) and $26.0\;\unit{mmHg}$ (venous pressure). Considering the average fluid pressure $p_{t,roi}$, we have mean values between $33.6\;\unit{mmHg}$ and $35.1\;\unit{mmHg}$ which means that there is potentially a flow from the arterial side through the tissue into the venous system. In Figure 9, one can see the four different stages of the growth process i.e. the initial configuration consisting of arterioles and venoles, the growth of the large scaled vessels, the larger vessels combined with the capillary bed and finally a slightly reduced network. For each case the oxygen distribution in a single tissue layer is depicted. In the first two phases, it is concentrated in a small vicinity of the vessels. After adding the capillary bed, we obtain a oxygen distribution covering almost the whole tissue domain, apart from the lower right corner which is not part of $\Omega_{roi}$.

\begin{table}[h!]
	\centering
	\caption{\label{tab:means_deviations} Mean values and standard deviations of the quantities of interest.}
	{\small
		\begin{center}
			\begin{tabular}{|c|c|c|c|c|}
				\hline
				\hline
				$\gamma\;\left[-\right]$	& $3.0$ & $3.5$ &  $3.0$ & $3.5$ \\
				\hline
				$m_0^{(ox)}\;\left[ \unitfrac{mmHg}{s} \right]$	& $3.0$ & $3.0$ &
				$4.0$ & $4.0$ \\
				\hline
				$L\;\left[\unit{m} \right] $ & $0.536 \pm 0.145$ &  $0.462 \pm 0.121$ & $0.751 \pm 0.159$ & $0.582 \pm 0.187$   \\
				\hline
				$A\;\left[\mu \unit{m}^2 \right] \cdot 10^{-5}$ & $10.5 \pm 2.49 $ & $9.38 \pm 2.10$  & $1.35 \pm 0.27$  & $1.15 \pm 0.32$ \\
				\hline
				$V\;\left[\unit{m}^3 \right]\cdot 10^{-11}$ & $1.88 \pm 0.33$ &  $ 1.78 \pm 0.29$ & $2.32 \pm 0.38$  
				& $2.09 \pm 0.45$ \\
				\hline
				$N_{seg}\;\left[\unit{-} \right]$ & $13845 \pm 850$ & $11689 \pm 685$  & $18459 \pm 845$  &  $15327 \pm 1241$ \\
				\hline
				$PO_{2,roi}^{(t)}\;\left[\unit{mmHg} \right]$ & $34.8 \pm 2.0$  &  $32.1 \pm 2.4$ &  $35.0 \pm 1.9$  &  $33.6 \pm 1.5$ \\
				\hline
				$p_{t,roi}\;\left[ \unit{mmHg} \right]$ & $34.8 \pm 2.1$ &  $34.0 \pm 2.4$ & $35.1 \pm 1.9$ & $33.6 \pm 1.5$ \\
				\hline
				$F_{tv}\;\left[ \unitfrac{\mu g}{s} \right] \cdot 10^{-3} $ & $8.22 \pm 1.28$ & $7.19 \pm 1.08$ 
				& $9.10 \pm 1.00$ & $8.24 \pm 1.10$ \\
				\hline
				$N_{it}\;\left[\unit{-} \right]$ & $34.2 \pm 2.2$ & $34.0 \pm 1.9$  & $36.1 \pm 2.1$  & $35.7 \pm 3.2$ \\
				\hline
				\hline
			\end{tabular}
		\end{center}
	}
\end{table} 

\section{Conclusion}
\label{sec:Conclusion}

In this paper, we have presented a model for simulating flow and oxygen transport processes in vascular systems embedded in a homogeneous tissue block. Using the distribution of the partial pressure of oxygen in 
tissue, a stochastic model for generating surrogate microvascular networks out of given arterioles and venoles has been constructed. Thereby, the gradient in the oxygen distribution governs the growth direction while by means of log-normal distributions the length of the vessels and the formation of bifurcations is determined. The choice of the vessel radii at a bifurcation as well as the branching angles are based on Murray's law. Combining characteristic distributions for lengths and radii with Murray's law and the oxygen gradient as the main growth direction, the resulting microvascular network incorporates several well-known optimality principles. In addition, the algorithm avoids an intersection of newly created vessels. Finally, it is guaranteed that in the interior part of the tissue domain no terminal vessels occur, i.e., the flow paths between the arterioles and venoles are not interrupted.

Our simulation results reveal that about $20$ simulations are needed to obtain a representative data set. Furthermore, we have observed that a Murray parameter of $\gamma=3$ and a low oxygen consumption rate yields a satisfactory match with respect to some characteristic quantities like the total surface area or the volume of the vascular network.

Future work in this field could be concerned with applying the presented numerical model to further similar data sets i.e. microvascular networks that are contained in a volume of a few cubic millimeters. For example it would be of great interest, how the values for the consumption rate of oxygen and the Murray parameter $\gamma$ have to be chosen to obtain satisfactory results. Moreover, one could try to find a relationship for the radii and the branching angles at a bifurcation that is more precise than Murray's law, which is only an approximation and does not hold in general. In addition to that, it could be investigated whether the remaining parameters like the mean values and standard deviations of the statistical distributions as well as the probability threshold for bifurcations $P_{th}$ have to be recalibrated. Finally, the performance of the numerical model could be further tested by considering CT scans showing the saturation distribution of a contrast fluid \cite{rohan2018modeling}. Besides the transport of oxygen further convection-diffusion equations modelling the injection of a contrast fluid could be added and the resulting simulation data could be compared with the image data.

\section*{Acknowledgements}
This work was partially supported by the DFG grant (WO/671 11-1). We gratefully acknowledge the authors of~\cite{reichold2009vascular} and, in particular, Prof. Dr. Bruno Weber and Prof. Dr. Patrick Jenny, for providing the data set of the vascular network considered in this paper.

\bibliography{Literature}
\nocite{*}
\bibliographystyle{plain}
	
\end{document}